%% file: main.tex
\documentclass[aps,pre,reprint,groupedaddress,floatfix]{revtex4-2}
\usepackage{xr}
\input{standardPkgs}
\newcommand{\tr}{\text{tr}}
\newcommand{\kB}{k_{\text{B}}}

\begin{document}

\title{Speed-Fisher Information: Chaos and Irreversibility in Classical and Quantum Dynamics}
\author{Nachiket Karve}
\email{nachiket@bu.edu}
\author{Nathan Rose}
\author{David Campbell}
\author{Anatoli Polkovnikov}
\affiliation{Department of Physics, Boston University, Boston, Massachusetts 02215, USA}
\date{\today}

\begin{abstract}
    We propose a unified perspective on classical and quantum chaos based on the sensitivity of a system's stationary states to slow driving. We probe this sensitivity via the system's susceptibility to the average protocol speed, which we call the ``speed-Fisher information," and relate it to irreversible entropy production in the system. We show that chaotic dynamics manifests as a divergence of the speed-Fisher information with the protocol time, and that this response is controlled by the perturbation's low-frequency spectral weight. This approach to chaos applies to both classical and quantum Hamiltonian systems, and naturally extends to non-Hamiltonian classical flows. Additionally, we identify a quantum regime at times shorter than the Planckian timescale, in which the speed-Fisher information exhibits a uniquely quantum scaling. We illustrate this framework with simple classical and quantum examples.
\end{abstract}

\maketitle

\section{Introduction}


Traditionally, chaos has been understood through the presence of long-time unpredictability in a system, despite the system's dynamics being governed by deterministic laws~\cite{ott_1993}. In classical systems, this unpredictability is commonly measured by the dynamical instability of phase-space trajectories~\cite{lorenz_1972,gleick_1987} and quantified by the presence of positive Lyapunov exponents~\cite{oseledets_1968}. However, trying to naively apply this definition to quantum systems fails, as states evolving unitarily under the same Hamiltonian preserve their inner product, and thus they do not separate exponentially ~\cite{berry_1989}. In systems with a well-defined classical limit, quantum chaos is commonly diagnosed through spectral signatures~\cite{bohigas_1984,berryTabor_1977} such as level repulsion and Wigner-Dyson statistics, consistent with random matrix theory~\cite{wigner_1951aa,dyson_1962aa,dyson_1962ab,dyson_1962ac}. 
More recently, the eigenstate thermalization hypothesis (ETH) has provided a complementary perspective, identifying quantum chaos with the structure of many-body eigenstates and their ability to encode thermal behavior~\cite{deutsch_1991,srednicki_1994,dalessio_2016}. Because these particular metrics for quantum and classical chaos are fundamentally distinct, they do not represent a comprehensive framework that seamlessly explains (i) the emergence of classical chaos from the underlying quantum dynamics and (ii) the emergence of long-term dynamical instabilities in quantum systems, especially close to the integrable, non-chaotic limits.

Recent work has established the adiabatic gauge potential (AGP)~\cite{kolodrubetz_2017}, the generator of adiabatic deformations, as a sensitive probe of both classical and quantum chaos~\cite{pandey_2020,kim_2026,karve_2025aa,pozsgay_2024,rose_2025,leblond_2020, vidmar_2025, orlov_2023, sharipov_2024, peres_1984}. These results suggest that the adiabatic variation of any parameter in a chaotic system, both quantum and classical, becomes increasingly complex. Building on these results, we propose that the sensitivity of stationary states to finite-speed driving provides an operational probe of chaos. The adiabatic theorem states that sufficiently slow Hamiltonian deformations preserve adiabatic invariants in classical systems and produce adiabatic following of instantaneous eigenstates in quantum systems~\cite{dirac_1925,arnold_1989,messiah_2014}. Because such adiabatic behavior can occur in both regular and chaotic systems, the strict adiabatic limit does not by itself distinguish between them. Their leading response at finite driving speed, however, can differ markedly. We therefore characterize chaos through the susceptibility of a stationary state to the driving speed of an applied perturbation.

\begin{figure}
    \centering
    \includegraphics[scale=1]{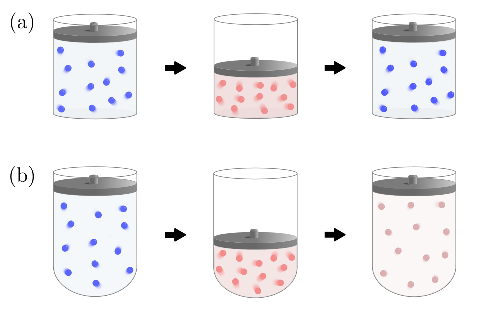}
    \caption{Slow, cyclic driving of (a) a regular system, whose final state closely approaches its initial state, and (b) a chaotic system, where thermodynamic friction produces dissipated work at finite driving speeds.}
    \label{fig_sketch}
\end{figure}

For concreteness, we consider stationary states subjected to slow cyclic driving. As we show in this work, chaotic dynamics can sustain a nonzero thermodynamic friction coefficient in the adiabatic limit (see Fig.~\ref{fig_sketch}). We call the corresponding susceptibility to changes in the speed of the drive the ``speed-Fisher information"~\cite{fisher_1922}. Fisher information is central to quantum information theory~\cite{wooters_1981,lambert_2023} and metrology~\cite{toth_2012,Liu_2020}, and has been widely used to detect quantum phase transitions~\cite{gu_2008,carollo_2020}. Here, instead of considering a manifold of parametrized eigenstates, the Fisher information is evaluated for final states generated by cyclic drives of different speeds. We show that this Fisher information is governed by the perturbation's low-frequency spectral weight. It remains nonsingular in regular systems, whereas, in thermalizing systems, perturbations that obey the ETH lead to a robust $\nicefrac{1}{\mu}$ divergence as the inverse protocol duration $\mu$ approaches zero. Furthermore, chaotic systems that exhibit nonthermalizing dynamics produce even stronger divergences.

We additionally show that the speed-Fisher information can be interpreted as a thermodynamic drag. This allows us to probe the speed-Fisher information by measuring the system's change in modular energy during the cyclic drive, which is equivalent to the dissipated work when the system is initially prepared in a thermal Gibbs state. This property makes the speed-Fisher information accessible both numerically and experimentally. This framework applies uniformly to quantum and classical systems, including non-Hamiltonian deterministic flows, as we demonstrate through representative examples.

In this work, we also study the speed-Fisher information outside the semiclassical, quasistatic regime in thermal many-body quantum systems. For observables with exponentially decaying correlations, we identify crossovers universally controlled by the ``Planckian time" $\nicefrac{\hbar}{\kB T}$~\cite{hartnoll_2022}, independent of system size. As the duration of the protocol is increased, the speed-Fisher information crosses from $\mathscr I_{\bar v}(\mu)\sim \nicefrac{1}{\mu^4}$ to eventually the classical scaling of $\mathscr I_{\bar v}(\mu)\sim \nicefrac{1}{\mu}$. This short-time $\nicefrac{1}{\mu^4}$ regime is absent classically and therefore represents a purely quantum contribution to the susceptibility.

This article accompanies Ref.~\cite{karve_2026}, where we present the main results in a compact Letter. Here, we provide detailed derivations for quantum, classical Hamiltonian, and non-Hamiltonian systems; establish the connection to entropy production and thermodynamic friction; analyze the many-body quantum regime; and present additional numerical examples. This paper is organized as follows. In Sec.~\ref{sec_sfi}, we introduce the speed-Fisher information and derive its general spectral representation. Sec.~\ref{sec_entropy} relates the abstract notion of the speed-Fisher information to numerically and experimentally accessible quantities, such as the change in modular energy of the system. In Sec.~\ref{sec_crossover}, we discuss the differences that can be observed in quantum and classical systems, and provide an estimate of the quantum-to-classical crossover timescale. Finally, in Sec.~\ref{sec_num}, we test our framework in a variety of numerical examples, including few- and many-body quantum and classical systems.

\section{Speed-Fisher Information}
\label{sec_sfi}
Fisher information is a well-established tool in information theory to quantify the sensitivity of a statistical distribution to changes in a parameter~\cite{lambert_2023,Liu_2020}. Consider a family of probability distributions $\rho(x|\theta)$ that depend on the external parameter $\theta$. The Fisher information quantifies how susceptible $\rho$ is to changes in this parameter and is defined as the variance of the score function~\cite{fisher_1922}:
\begin{equation}
    \mathscr{I}_\theta = \int dx \ \rho(x|\theta) \left(\frac{\partial \ln \rho(x|\theta)}{\partial\theta}\right)^2.
\end{equation}
A large Fisher information indicates sensitive dependence on the parameter $\theta$. Equivalently, it determines how distinguishable neighboring distributions $\rho(x|\theta)$ and $\rho(x|\theta+\delta\theta)$ are.


A family of distributions commonly used to describe physical states is a set of stationary distributions generated by a static, parameter-dependent Hamiltonian. Of course, this is not the only possible family of distributions over which one can apply Fisher information. Instead, we consider a family of distributions generated by dynamical evolution under a set of time-dependent Hamiltonians with parameter-dependent protocols. Cyclic deformations provide a particularly natural setting for defining such a dynamical susceptibility. Consider a system described by the Hamiltonian $H_0$, initially prepared in a stationary state $\rho_{-\infty}$ and subjected to a time-dependent deformation
\begin{equation}
H(\tau)=H_0+\lambda(\tau)V, \quad \lambda(\pm\infty)=0.
\label{eq_timeDepHam}
\end{equation}
Because the Hamiltonian returns to $H_0$ at the end of the cycle, the final state can be compared directly with the initial state. In an ideal reversible quasistatic cycle, the state follows the instantaneous stationary states and returns to the same statistical state upon completion of the cycle. However, the system need not follow this quasistatic evolution exactly at a nonzero driving speed. This motivates us to treat the driving speed itself as the parameter with respect to which the susceptibility is evaluated. We will refer to this susceptibility as the ``speed-Fisher information''. Since chaos is generally associated with an enhanced sensitivity to perturbations, one might expect chaotic systems to exhibit large speed-Fisher information under generic cyclic driving. In the rest of this section, we study speed-Fisher information in quantum, classical, and classical non-Hamiltonian systems, and discuss its relation to the low-frequency spectral content of the perturbation. Finer details of the calculations are provided in Appendices~\ref{app_qsfi}, \ref{app_csfi}, and \ref{app_nhsfi}.

\subsection{Quantum Systems}

Consider a quantum system prepared in a stationary state described by the density matrix $\rho_{-\infty}$ and evolved under the time-dependent Hamiltonian in Eq.~(\ref{eq_timeDepHam}). Stationarity of the state under unperturbed dynamics implies $[H_0, \rho_{-\infty}] = 0$. In the absence of additional conserved quantities, it is natural to take this initial stationary state to be a function of the Hamiltonian alone; that is,
\begin{equation}
    \rho_{-\infty} = P(H_0).
\end{equation}
In integrable systems, one can have more general stationary distributions described by Generalized Gibbs ensembles 
\cite{vidmar_2016}, but these are usually harder to realize in generic preparation schemes, since coupling such systems to an environment typically yields equilibrium thermal ensembles~\cite{lange_2017}. While the assumption that $\rho_{-\infty}=P(H_0)$ is not essential in our derivations, for concreteness, we will restrict the analysis to such initial stationary states. In what follows, we assume that the function $P$ is smooth and sufficiently well-behaved.

The perturbation $\lambda(\tau)$ generally depends on two parameters, namely, the maximum amplitude $\lambda_{\text{max}}$ and the duration of the protocol $t$. Since we are interested in the susceptibility of the final state to the driving speed, it is convenient to trade the amplitude for the mean speed of the protocol. We therefore parameterize the perturbation as
\begin{equation}
    \lambda(\tau) = \frac{\bar{v}}{\mu} f(\mu \tau),
\end{equation}
along with the condition that
\begin{equation}
    \int_{-\infty}^{+\infty} dx \ |g(x)| = 1, \text{ where } g(x) = f'(x).
\end{equation}
Thus, the protocol is now parameterized by the inverse duration $\mu = \nicefrac{1}{t}$ and the average speed
\begin{equation}
    \bar{v} = \frac{1}{t} \int_{-\infty}^{+\infty} d\tau \ \left|\dot{\lambda}(\tau)\right|.
\end{equation}
For a fixed protocol shape, $(\lambda_{\text{max}}, t)$ and $(\bar{v},\mu)$ are equivalent parameterizations.

The final state of the system $\rho_{+\infty}$ after the completion of the protocol can be estimated using first-order perturbation theory. For general protocols, this result is only valid as long as $\lambda_{\text{max}} \ll 1$, or equivalently when $\bar{v} \ll \mu$ in natural units. However, if $\lambda$ represents some cyclic variable, such as an angle of rotation of a magnetic or electric field or a phase of a periodic function, then there is no restriction on $\lambda_{\rm max}$. Moreover, in this case, $\lambda(t)$ does not need to return to the initial value as long as the Hamiltonian of the system is the same at the end of the protocol as at the beginning. Working in the interaction picture with respect to $H_0$, the Dyson expansion at a fixed $\mu$ allows us to compute the final state of the system to leading order in $\bar{v}$. We measure the overlap between the initial and final states $\rho_{\pm\infty}$ through the fidelity~\cite{jozsa_1994,nielsen_2010}, which depends on both the speed and duration of the protocol, and thus, we denote it as $\mathscr{F}(\bar{v},\mu)$. In quantum systems, the fidelity is defined as~\cite{bures_1969,helstrom_1967}
\begin{equation}
    \mathscr{F}(\bar{v},\mu) = \left(\tr\sqrt{\sqrt{\rho_{-\infty}}\ \rho_{+\infty} \ \sqrt{\rho_{-\infty}}}\right)^2.
\end{equation}
It obeys $\mathscr{F}(\bar{v},\mu) \leq 1$, with equality if and only if the final state coincides with the initial state. Importantly, as shown in Appendix~\ref{app_qsfi}, up to leading order in $\bar{v}$ this fidelity depends on the symmetric auto-correlation function of $L_V$, defined as
\begin{equation}
    C_{L_V}(\tau_1-\tau_2) = \frac{1}{2}\langle\{L_V(\tau_1),L_V(\tau_2)\}_+ \rangle,
\end{equation}
where $\{ \ , \ \}_+$ denotes an anti-commutator, $\langle \ \rangle$ denotes an average over $\rho_{-\infty}$, and the symmetric logarithmic derivative $L_V$ is defined implicitly through
\begin{equation}
    \frac{1}{2}\left(\rho_{-\infty} L_V(\tau) + L_V(\tau)\rho_{-\infty}\right) = \frac{1}{i\hbar}[V(\tau),\rho_{-\infty}],
\end{equation}
with $V(\tau) = e^{iH_0\tau/\hbar}Ve^{-iH_0\tau/\hbar}$ denoting Heisenberg evolution under $H_0$. Then, the fidelity can be expressed as
\begin{equation}
    \mathscr{F}(\bar{v},\mu) = 1 - \frac{\bar{v}^2}{4\mu^3} \int_{-\infty}^\infty \frac{d\omega}{2\pi} \ \tilde{C}_{L_V}(\mu \omega) \left|\tilde{f}\left(\omega\right)\right|^2 + \mathcal O(\bar v^3).
    \label{eq_fidelity}
\end{equation}
Here, a tilde denotes a Fourier transform, with the convention
\begin{equation}
    \tilde{h}(\omega) = \int_{-\infty}^\infty d\tau \ e^{-i\omega \tau} h(\tau).
\end{equation}
From here on, we will refer to $\tilde{C}_{L_V}$ as the spectral function of $L_V$.

The leading correction to the fidelity in Eq.~(\ref{eq_fidelity}) is of order $\bar{v}^2$. Thus, the speed-Fisher information, which is defined as the fidelity susceptibility with respect to mean protocol speed at a finite $\mu$, is given by
\begin{align}
    \mathscr{I}_{\bar{v}}(\mu) &= -2\frac{\partial^2\mathscr{F}}{\partial \bar{v}^2}\Biggr|_{\bar{v}=0} \notag\\&= \frac{1}{\mu^3} \int_{-\infty}^\infty \frac{d\omega}{2\pi} \ \tilde{C}_{L_V}(\mu \omega) \left|\tilde{f}\left(\omega\right)\right|^2.
    \label{eq_fisherInf}
\end{align} 
The behavior of the speed-Fisher information as $\mu\to 0$ determines whether this susceptibility remains finite or becomes singular in the quasistatic limit. For a fixed protocol shape, apart from the explicit pre-factor $\nicefrac{1}{\mu^3}$, the dependence on $\mu$ enters only through $\tilde{C}_{L_V}(\mu\omega)$. The quasistatic scaling of the speed-Fisher information is therefore governed by the low-frequency spectral content of $L_V$.

To connect this result more directly to the physical perturbation $V$, we now use the assumption that $\rho_{-\infty}=P(H_0)$, with $P$ being a smooth function of energy. According to Eq.~(\ref{eq_fisherInf}), any singular dependence of the speed-Fisher information on $\mu$ in the limit $\mu\to 0$ is controlled by the low-frequency behavior of $\tilde{C}_{L_V}(\omega)$. Thus, as shown in Appendix~\ref{app_qsfi}, the spectral function $\tilde{C}_{L_V}(\omega)$ can be approximated by $\omega^2 \tilde{\mathcal{D}}_V(\omega)$, where $\tilde{\mathcal{D}}_V(\omega)$ is the Fourier transform of the ``score-weighted autocorrelation function" of $V$, defined as
\begin{equation}
    \mathcal D_V(\tau) = \frac{1}{2} \left\langle s(H_0)^2 \{\delta V(\tau),\delta V(0)\}_+ \right\rangle,
    \label{eq_DVt}
\end{equation}
with corrections vanishing in the limit $\hbar\omega\rightarrow 0$. Here $\delta V(\tau) = V(\tau) - \bar{V}$ is the perturbation with its infinite time-average removed, and the score function is defined as
\begin{equation}
    s(H_0) = \frac{\partial \ln \rho_{-\infty}}{\partial E}\bigg|_{E=H_0}\equiv  \frac{\partial \ln P(E)}{\partial E}\bigg|_{E=H_0}\,.
\end{equation}
Importantly, this approximation holds provided
\begin{equation}
\hbar\mu
\left|\partial_E\ln P(E)\right|
\ll 1
\label{eq_semiClassCond}
\end{equation}
for all energies appreciably supported by $P(E)$~\footnote{More precisely, the low-frequency contribution is sampled at $\omega_{mn}=\mu x$, with $x$ in the range appreciably weighted by the driving protocol. In regular systems, the leading finite contribution to the Fisher information may instead originate from spectral features at frequencies that are independent of $\mu$. Nevertheless, the low-frequency approximation above correctly establishes that this regime does not generate a divergence, although it need not reproduce the finite regular background quantitatively.}. Finally, the asymptotic relation $\tilde{C}_{L_V}(\omega)\approx \omega^2\tilde{\mathcal{D}}_V(\omega)$ shows that any zero-frequency delta-function contribution (Drude weight) to $\tilde{\mathcal{D}}_V(\omega)$ can be removed without affecting the speed-Fisher information, since $\omega^2\delta(\omega)=0$.

Finally, the speed-Fisher information can be expressed in terms of the score-weighted spectral function of $V$, as
\begin{equation}
    \mathscr{I}_{\bar{v}}(\mu) \simeq \frac{1}{\mu}\int_{-\infty}^\infty \frac{d\omega}{2\pi} \ \tilde{\mathcal{D}}_V(\mu \omega) \left|\tilde{g}(\omega)\right|^2.
    \label{eq_qsfi}
\end{equation}
In the above expression, the shape of the protocol enters only through the filter function $|\tilde g(\omega)|^2$. The quasistatic scaling is therefore controlled by a combination of the low-frequency behavior of the score-weighted spectral function and the high-frequency tail of the protocol filter.

In this work, we focus on the two most common stationary states: 

\begin{itemize}
\item[i)] The Gibbs ensemble with $\rho_{-\infty} = \frac{1}{Z}e^{-H_0/\kB T}$, where the score is simply $\nicefrac{-1}{\kB T}$. Hence, the score-weighted correlation function is equivalent to the standard one:
\begin{equation}
    \mathcal{D}_V(\tau) = \frac{(\kB T)^{-2}}{2}\left\langle \ [\delta V(\tau) , \delta V(0)]_+ \right\rangle.
\end{equation}
Here, the condition in Eq.~(\ref{eq_semiClassCond}) becomes
\begin{equation}
    \frac{\hbar\mu}{\kB T} \ll 1.
\end{equation}
This includes the semiclassical limit $\hbar\to 0$, the quasistatic limit $\mu\to 0$, and the high-temperature limit $T\to \infty$.

\item[ii)] The Gaussian microcanonical ensemble localized within a narrow energy window; that is, $\rho_{-\infty} \propto e^{-(H_0-E)^2/2\sigma^2}$. In this case,  $s(H_0)=-\tfrac{1}{\sigma^2}(H_0-E)$, and thus the score-weighted correlation function is given by
\begin{equation}
    \mathcal{D}_V(\tau) = \frac{1}{2\sigma^4}\left\langle (H_0-E)^2 \ [\delta V(\tau) , \delta V(0)]_+ \right\rangle.
\end{equation}
In this case, Eq.~(\ref{eq_semiClassCond}) becomes
\begin{equation}
    \frac{\hbar \mu}{\sigma} \ll 1.
\end{equation}

\end{itemize}

\subsection{Classical Hamiltonian Systems}

The analysis of the speed-Fisher information in classical systems follows arguments similar to those in the quantum case. The system is now initialized in a stationary state described by the probability distribution $\rho_{-\infty}(\mathbf{x})$, where $\mathbf{x}$ denotes the system's phase-space coordinates. After adding a perturbation $\frac{\bar{v}}{\mu} f(\mu \tau) V(\mathbf{x})$, the final state of the system is described by the distribution $\rho_{+\infty}(\mathbf{x})$. The classical fidelity between the initial and final states is given by~\cite{hellinger_1909}
\begin{equation}
    \mathscr{F}(\bar{v},\mu) = \left(\int d\mathbf{x} \sqrt{\rho_{+\infty}(\mathbf{x})\rho_{-\infty}(\mathbf{x})}\right)^2.
    \label{eq_classicalFid}
\end{equation}
Like the quantum version, the fidelity can be expressed in terms of the spectral function of $L_V$, where
\begin{equation}
    C_{L_V}(\tau) = \langle L_V(\tau) L_V(0) \rangle
\end{equation}
and $L_V(\tau) = \{V(\mathbf{x}(\tau)),\ln\rho_{-\infty}(\mathbf{x}(\tau))\}$ is the classical logarithmic derivative. We find that
\begin{equation}
    \mathscr{I}_{\bar{v}}(\mu) = \frac{1}{\mu^3} \int_{-\infty}^\infty \frac{d\omega}{2\pi} \ \tilde{C}_{L_V}(\mu \omega) \left|\tilde{f}\left(\omega\right)\right|^2.
\end{equation}
Furthermore, assuming $\rho_{-\infty}(\mathbf{x}) = P(H_0(\mathbf{x}))$ to be a smooth function of $H_0$, the spectral function of $L_V$ can be expressed in terms of the score-weighted spectral function of $V$ as
\begin{equation}
    \tilde{C}_{L_V}(\omega) = \omega^2 \tilde{\mathcal{D}}_V(\omega),
\end{equation}
with $\mathcal{D}_V(\tau)$ given by the classical version of Eq.~(\ref{eq_DVt}):
\begin{align}
    \mathcal{D}_V(\tau) =& \ \int d\mathbf{x} \ P(E(\mathbf{x})) \ s(E(\mathbf{x}))^2 \ \delta V(\mathbf{x}(\tau)) \ \delta V(\mathbf{x}(0)).
\end{align}
The classical speed–Fisher information therefore takes the exact form:
\begin{equation}
    \mathscr{I}_{\bar{v}}(\mu) = \frac{1}{\mu} \int_{-\infty}^\infty \frac{d\omega}{2\pi} \ \tilde{\mathcal{D}}_{V}(\mu \omega) \left|\tilde{g}\left(\omega\right)\right|^2.
    \label{eq_csfi}
\end{equation}
Eqs.~(\ref{eq_qsfi}) and (\ref{eq_csfi}) have the same functional form, but the former is a low-frequency quantum approximation while the latter is an exact classical identity.

\subsection{Classical Autonomous Systems}

We now consider the case of an autonomous system, where the governing differential equation of the unperturbed system is of the form
\begin{equation}
    \frac{d\mathbf{x}}{d\tau} = \mathbf{F}(\mathbf{x}).
\end{equation}
Here, $\mathbf{x}$ is a multi-dimensional vector that encodes the state of the system. We introduce a perturbation, so that the differential equation takes the form $\frac{d\mathbf{x}}{d\tau} = \mathbf{F}(\mathbf{x}) + \lambda(\tau)\mathbf{V}(\mathbf{x})$. Note that $\mathbf{F}$ represents a velocity field; therefore, the speed of the perturbation $\bar{v}$ is captured by the strength of the perturbation itself, and we parameterize $\lambda$ as
\begin{equation}
   \lambda(\tau) =  \bar{v} f(\mu \tau), \text{ with } \int_{-\infty}^\infty |f(z)| \ dz = 1.
\end{equation}

As before, we initialize the system in $\rho_{-\infty}(\mathbf{x})$, which is a stationary probability distribution of $\mathbf{F}$, and let the system evolve over time. We assume that $\rho_{-\infty}$ is smooth and positive on its support and that boundary terms vanish. Systems with singular invariant measures require an appropriate coarse-graining or regularization. 
Like the classical Hamiltonian case, the fidelity between the initial and final states is given by Eq.~(\ref{eq_classicalFid}), and the speed-Fisher information can be shown to be
\begin{equation}
    \mathscr{I}_{\bar{v}}(\mu) = \frac{1}{\mu}\int_{-\infty}^\infty \frac{d\omega}{2\pi} \ \tilde{C}_{L_V}(\mu \omega)|\tilde{f}(\omega)|^2,
\end{equation}
where the logarithmic derivative is now defined as
\begin{equation}
    L_V(\mathbf{x}) = \frac{1}{\rho_{-\infty}(\mathbf{x})}\nabla \cdot [\rho_{-\infty}(\mathbf{x}) \mathbf{V}(\mathbf{x})].
    \label{eq_logDerNonHam}
\end{equation}
The logarithmic derivative is therefore the divergence of the perturbing flow with respect to the stationary measure. Positive values of $L_V$ describe a local outflow of probability under the perturbation, while negative values describe a local influx. The logarithmic derivative thus directly measures the local deformation of the stationary distribution generated by $\mathbf V$. Thus, our observable of interest in this case is the logarithmic derivative itself, and not the perturbation. Note that the expression for the speed-Fisher information is analogous to Eqs.~(\ref{eq_qsfi}) and (\ref{eq_csfi}). Once again, the low-frequency behavior of the spectral function $\tilde{C}_{L_V}$ determines the scaling of the speed-Fisher information with $\mu$.

\subsection{Asymptotic Behavior of the Speed-Fisher Information}

\begin{table}
\begin{ruledtabular}
    \vspace*{1em}
    \begin{tabular}{lll}
        & $\tilde{\mathcal{D}}_{V}(\omega\to 0)$ & $\mathscr{I}_{\bar{v}}(\mu\to 0)$ \\
        \midrule
        Regular & $\sim |\omega|^{\alpha} \ (\alpha > 1)$ & $\mathcal O\left(\mu^0\right)$ \\
        \midrule
        \makecell[l]{Marginally \\ unstable} & $\sim |\omega|^{\alpha}\ (0<\alpha \leq 1)$ & $\mathcal{O}\left({\nicefrac{1}{\mu^{1\!-\!\alpha}}}\right)$ \\
        \midrule
        \makecell[l]{Chaotic \\ thermalizing} & $\text{const} > 0$ & $\mathcal{O}\left({\nicefrac{1}{\mu}}\right)$ \\
        \midrule
        \makecell[l]{Chaotic \\ nonthermalizing} & $\sim |\omega|^{\alpha} \ (\alpha < 0)$ & $\mathcal{O}\left({\nicefrac{1}{\mu^{1\!+\!|\alpha|}}}\right)$ \\
    \end{tabular}
\end{ruledtabular}
\caption{Relation between the low-frequency spectral weight and the speed-Fisher information. For the regular regime and a smooth Gaussian-type filter function $f(z)$, $\mathscr{I}_{\bar{v}}(\mu\to 0)\sim \mu^{\alpha-1}$}
\label{tab_chaos}    
\end{table}

Having derived the relation between the speed-Fisher information and the low-frequency spectral content of the perturbation, we now analyze its behavior in different regimes. In particular, we discuss the expected scaling of the speed-Fisher information with $\mu$ under regular and chaotic dynamics in this section. See Table~\ref{tab_chaos} for a summary of the different regimes.

For regular or integrable dynamics, the spectral functions of observables corresponding to integrability-preserving perturbations or weak integrability-breaking perturbations~\cite{surace_2023,vanovac_2024,vanovac_2026} vanish at low frequency, at least linearly and typically quadratically. This condition guarantees perturbative stability of classical orbits/quantum eigenstates~\cite{kim_2026}. Thus, it is reasonable to assume that
\begin{equation}
    \exists \delta > 0 \text{ such that for } |\omega| < \delta, \ |\tilde{\mathcal{D}}_V(\omega)| \leq C_1 |\omega|^\alpha,
    \label{eq_specFnCond}
\end{equation}
for some $C_1 > 0$ and $\alpha > 1$. Moreover, we assume that $\tilde{g}(\omega)$ decays sufficiently fast, that is
\begin{equation}
    |\tilde{g}(\omega)| \leq \frac{C_2}{|\omega|},
    \label{eq_gDecay}
\end{equation}
 for some $C_2 > 0$. Under these conditions, the speed-Fisher information does not diverge in the quasistatic limit: depending on the protocol shape, it either vanishes with $\mu$ or saturates to a finite value. Physically, the finite positive value of the speed-Fisher information for regular motion in the limit $\mu\to 0$ appears due to transient excitations. For smooth protocols, increasing the protocol time $\nicefrac{1}{\mu}$ eliminates transients and $\mathscr{I}_{\bar{v}}(\mu\to 0)\to 0$. For protocols in which the velocity has a sudden jump, $\mathscr{I}_{\bar{v}}(\mu\to 0)$ generally saturates at a positive constant independent of $\mu$.
 
By contrast, chaotic systems have a non-vanishing low-frequency spectral weight, and often satisfy the low-frequency asymptotic scaling for the spectral function given by Eq.~(\ref{eq_specFnCond}), with the condition that $-1 < \alpha \leq 0$. Then, the speed-Fisher information scales as $1/\mu^{1+|\alpha|}$ in the quasistatic limit. Generic perturbations in chaotic thermalizing systems, i.e., systems which satisfy the ETH~\cite{deutsch_1991,srednicki_1994,dalessio_2016}, exhibit a low-frequency spectral plateau ($\alpha=0$), producing a robust $\nicefrac{1}{\mu}$ divergence of $\mathscr I_{\bar v}$. On the other hand, nonthermalizing chaotic systems can have singular low-frequency spectra, see Refs.~\cite{leblond_2020,kim_2026}, leading to an even stronger divergence. 
 
Finally, spectral functions that satisfy the ansatz in Eq.~(\ref{eq_specFnCond}) with $0 < \alpha < 1$ are found in critical~\cite{pereira_2008}, or sub-Ohmic systems~\cite{leggett_1987}. In this case, the speed-Fisher information exhibits a weaker divergence of $1/\mu^{1-|\alpha|}$.

\subsection{Relation to the Regularized AGP Norm}

\begin{figure*}[htb]
    \centering
    \includegraphics[]{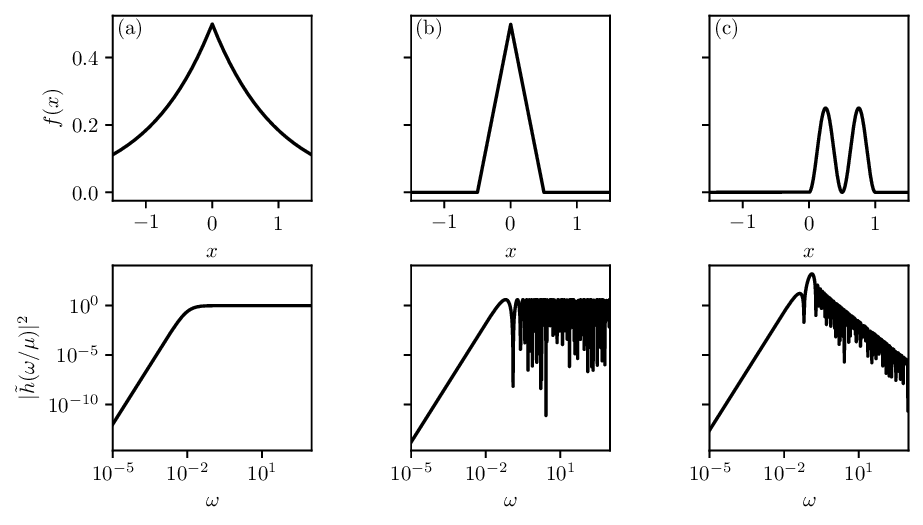}
    \begin{subfigure}[t]{\linewidth}
        \phantomcaption
    \label{fig_pertA}    
    \end{subfigure}
    \begin{subfigure}[t]{\linewidth}
        \phantomcaption
    \label{fig_pertB}    
    \end{subfigure}
    \begin{subfigure}[t]{\linewidth}
        \phantomcaption
    \label{fig_pertC}    
    \end{subfigure}
    \caption{The perturbation profile (top) and the corresponding acceleration filter (bottom) for three different protocols: (a) exponentially modulated, (b) triangular, and (c) sinusoidal. Parameters used -- $\mu = 10^{-2}$. The high-frequency noisy profile is an artifact of the logarithmic scale.}
\end{figure*}

We now discuss how the speed-Fisher information, under an appropriate protocol, can be used to regularize and measure the corresponding AGP norm. Recall that the AGP is defined as the generator of deformations under slow variations of a specific parameter in the system~\cite{kolodrubetz_2017}. Thus, the AGP norm gives the regularized fidelity susceptibility associated with a static deformation of the Hamiltonian~\cite{delcampo_2012,delcampo_2017,delcampo_2024}. In quantum systems, the AGP operator is defined as
\begin{equation}
\mathcal{A}_\lambda \ket{n(\lambda)} = i\hbar\partial_\lambda \ket{n(\lambda)},
\end{equation}
where the $\ket{n(\lambda)}$ are the instantaneous eigenstates of the system described by the Hamiltonian $H(\lambda) = H_0 + \lambda V$, and $\lambda$ is the parameter being tuned. The corresponding AGP in a classical system is defined through a Wigner-Weyl transform, and becomes a phase-space function of the canonical variables. Physically, it is the generator of trajectory-preserving canonical transformations~\cite{kim_2026}. In both quantum and classical systems, it is well established that the norm of the AGP, $||\mathcal{A}_\lambda||^2$, serves as a sensitive probe of chaos. Crucially, this norm is related to the spectral function of the perturbation $V$ as
\begin{equation}
    ||\mathcal{A}_\lambda||^2 = \int_{-\infty}^{\infty} \frac{d\omega}{2\pi} \ \frac{\tilde{C}_V(\omega)}{\omega^2}.
\end{equation}
In instances where the spectral function of the perturbation contains significant low-frequency weight, i.e. chaotic systems, the above integral diverges due to the $\nicefrac{1}{\omega^2}$ factor blowing up as $\omega\to 0$. To mitigate this issue, one must introduce a regularizer $\mu$ that suppresses contributions from frequencies smaller than $\mu$, while contributions from larger frequencies are kept intact. That is, the AGP norm is now defined as a function of the regularizer $\mu$ as
\begin{equation}
    ||\mathcal{A}_\lambda(\mu)||^2 = \int_{-\infty}^{\infty} \frac{d\omega}{2\pi} \ \frac{\tilde{C}_V(\omega)}{\omega^2} \left|\tilde{h}\left(\frac{\omega}{\mu}\right)\right|^2,
\end{equation}
where the filter function $\tilde{h}(\omega)$ must preserve large-frequency contributions, i.e. 
\begin{equation}
    \lim_{\omega\to\infty} |\tilde{h}(\omega)| \neq 0,
    \label{eq_accCond}
\end{equation} 
and suppress the $\nicefrac{1}{\omega^2}$ factor at smaller frequencies, i.e. 
\begin{equation}
    |\tilde{h}(\omega\to 0)| \sim \mathcal{O}(\omega).
\end{equation}
Note the similarities between the expression for the AGP norm and the expression for the speed-Fisher information (Eqs.~(\ref{eq_qsfi}) and (\ref{eq_csfi})). The equivalence can be made more apparent by defining $h(x) = f''(x)$ to be the acceleration profile of the perturbation, and writing the Fisher information as
\begin{equation}
    \mathscr{I}_{\bar{v}}(\mu) = \int_{-\infty}^\infty \frac{d\omega}{2\pi} \ \frac{\tilde{\mathcal{D}}_{V}(\omega)}{\omega^2} \left|\tilde{h}\left(\frac{\omega}{\mu}\right)\right|^2.
\end{equation}
For a Gibbs distribution, $\tilde{\mathcal{D}}_V(\omega) = (\kB T)^{-2}\tilde{C}_V(\omega)$, and therefore, the speed-Fisher information and the AGP norm of $V$ are identical, up to a temperature-dependent factor, with the regularization provided by the acceleration filter $\tilde{h}(\omega)$. It is important to note that since $\tilde{h}(\omega)$ must obey Eq.~(\ref{eq_accCond}), the function $h(x)$ cannot be absolutely integrable on the real line (c.f. the Riemann--Lebesgue lemma~\cite{bochner_1949}). Thus, to correctly extract the AGP norm from the speed-Fisher information, the acceleration profile must necessarily be impulsive, i.e., have a discontinuity in the velocity. Below, we provide examples of such impulsive protocols. 

We first consider an exponentially modulated protocol of the form
\begin{equation}
    \lambda(\tau) = \frac{\bar v}{2\mu} e^{-\mu|\tau|},
    \label{eq_expProtocol}
\end{equation}
with a characteristic duration of $\nicefrac{1}{\mu}$ and average speed of $\bar{v}$. Here, the acceleration profile is
\begin{equation}
    \ddot\lambda(\tau) = \frac{\bar{v}}{2} e^{-\mu|\tau|}\left[\mu - 2\delta(\tau)\right],
\end{equation}
which has an impulse at $\tau=0$. Moreover, the acceleration filter of this protocol suppresses small frequencies while preserving large-frequency contributions, as required (see Fig.~\ref{fig_pertA}). This translates to the speed-Fisher information being
\begin{equation}
    \mathscr{I}_{\bar{v}}(\mu) = \int_{-\infty}^\infty \frac{d\omega}{2\pi} \ \frac{\omega^2}{(\mu^2+\omega^2)^2} \tilde{\mathcal{D}}_{V}(\omega).
\end{equation}
The above expression is identical to the exponentially regularized AGP norm (cf. Eq. (25) in Ref.~\cite{kim_2026}).

Another example of a protocol that allows us to extract the AGP norm is the triangular protocol (Fig.~\ref{fig_pertB}):
\begin{equation}
\label{eq:triangular_protocol}
    \lambda(\tau) = \begin{cases}\displaystyle\bar v\left(\frac{1}{2\mu} - |\tau|\right), & |\tau| \leq \displaystyle\frac{1}{2\mu}, \\
    0, & \text{otherwise,}
    \end{cases}
\end{equation}
which has the acceleration profile
\begin{equation}
    \ddot\lambda(\tau) = \bar{v} \left[\delta\left(\tau + \frac{1}{2\mu}\right) + \delta\left(\tau - \frac{1}{2\mu}\right) - 2\delta(\tau)\right],
\end{equation}
and corresponds to the Fisher information
\begin{equation}
    \mathscr{I}_{\bar{v}}(\mu) = 4\int_{-\infty}^\infty \frac{d\omega}{2\pi} \ \frac{\left[1 - \cos\left(\frac{\omega}{2\mu}\right)\right]^2}{\omega^2} \tilde{\mathcal{D}}_{V}( \omega).
\end{equation}
This expression is identical, up to a factor, to the AGP norm that is regularized by a finite-time cutoff.

On the other hand, one cannot extract the AGP norm from a protocol with an absolutely integrable acceleration profile. As an example, we consider a sinusoidal protocol that takes the form
\begin{equation}
    \lambda(\tau) =
    \begin{cases}
        \dfrac{\bar{v}}{4\mu}\sin^2(2\pi\mu\tau), & 0\leq \tau\leq \dfrac{1}{\mu},\\
        0, & \text{otherwise},
    \end{cases}
    \label{eq_drive}
\end{equation}
leading to
\begin{equation}
    \ddot\lambda(\tau) =
    \begin{cases}
        2\pi^2\bar{v}\mu \cos(4\pi\mu \tau), & 0\leq \tau\leq \dfrac{1}{\mu},\\
        0, & \text{otherwise}.
    \end{cases}
\end{equation}
In this case, the speed-Fisher information, according to Eqs.~(\ref{eq_qsfi}) and (\ref{eq_csfi}) can be written as
\begin{equation}
    \mathscr{I}_{\bar{v}}(\mu) = \frac{\mu^2}{16} \int_{-\infty}^\infty \frac{d\omega}{2\pi} \  \frac{\sin^2\left(\frac{\omega}{2\mu}\right)}{\left(\mu^2-\frac{\omega^2}{16\pi^2}\right)^2} \tilde{\mathcal{D}}_{V}(\omega).
\end{equation}
Note that this expression does not reproduce the unregularized AGP norm in the limit $\mu\to 0$. Moreover, its acceleration profile suppresses both small- and large-frequency contributions to the speed-Fisher information, as can be seen in Fig.~\ref{fig_pertC}.  Consequently, in regular systems, the speed-Fisher information vanishes in the limit $\mu\to 0$. In contrast, the AGP norm saturates at a positive constant as long as the perturbation does not commute with the Hamiltonian. At the same time, the speed-Fisher information for this protocol in the chaotic regime has the same asymptotic scaling as the AGP norm for sufficiently small $\mu$. We shall use this sinusoidal protocol throughout the numerics performed in Sec.~\ref{sec_num}.


\section{Irreversible Entropy Production and Thermodynamic Friction}
\label{sec_entropy}
Thus far, we have characterized the response to slow cyclic driving in information-geometric terms by quantifying the distinguishability between the initial and final states. We now connect this susceptibility to an experimentally accessible thermodynamic quantity. For Hamiltonian systems, the irreversibility generated by a cyclic protocol can be quantified by the entropy production, or equivalently, by the dissipated work for thermal states. As we show below, in the slow-driving regime, this entropy production is governed by the same dynamical correlations that determine the speed-Fisher information. This relation provides an operational route to extracting the speed-Fisher information without requiring full knowledge of the state and, through it, the regularized AGP norm and the components of the Fubini-Study metric tensor.

The entropy produced by the external drive can be computed from the relative entropy between the initial and final states. This is given by the Kullback–Leibler (KL) divergence~\cite{kullback_1951,umegaki_1962}:
\begin{equation}
    D_{\text{KL}}(\rho_{+\infty} || \rho_{-\infty}) = \tr\left\{\rho_{+\infty}\ln\rho_{+\infty} - \rho_{+\infty}\ln\rho_{-\infty}\right\}
    \label{eq_dkl_defnQuantum}
\end{equation}
in quantum systems, and
\begin{equation}
    D_{\text{KL}}(\rho_{+\infty} || \rho_{-\infty}) = \int d\mathbf{x} \ \rho_{+\infty}(\mathbf{x})\ln\left(\frac{\rho_{+\infty}(\mathbf{x})}{\rho_{-\infty}(\mathbf{x})}\right)
    \label{eq_dkl_defnClassical}
\end{equation}
in classical systems. As shown in Appendix~\ref{app_entropy}, in the quasistatic limit, the KL-divergence can be expressed in terms of the speed-Fisher information up to leading order in $\bar{v}$:
\begin{equation}
    D_{\text{KL}}(\rho_{+\infty} || \rho_{-\infty}) \simeq \frac{1}{2}\mathscr{I}_{\bar{v}}(\mu)\bar{v}^2.
    \label{eq_dkl}
\end{equation}
Moreover, if we define the ``modular Hamiltonian" as $K_0 = -\ln\rho_{-\infty}$, then the above expression allows us to compute the average change in modular energy under the cyclic protocol:
\begin{equation}
    \langle\Delta K_0\rangle \simeq \frac{1}{2}\mathscr{I}_{\bar{v}}(\mu)\bar{v}^2.
    \label{eq_modEnergyChange}
\end{equation}

Importantly, for the Gibbs ensemble, the modular Hamiltonian is simply $K_0 = \frac{1}{\kB T} H_0 + \ln Z$, and we find that the average energy gained by the system is given by
\begin{equation}
    \langle\Delta E\rangle \simeq \frac{\kB T}{2}\mathscr{I}_{\bar{v}}(\mu)\bar{v}^2.
\end{equation}
This allows us to interpret the speed-Fisher information as a thermodynamic drag and to extract it from the dissipated work in experiments. Since the duration of the driving protocol is $\nicefrac{1}{\mu}$, the average power is given by $\mu \langle\Delta E\rangle$. Thus, for a triangular protocol~\eqref{eq:triangular_protocol} of total duration $1/\mu$  
\begin{equation}
\label{eq:friction_coef}
    \zeta_\text{eff}(\mu)=\tfrac{\mu \kB T}2 \mathscr{I}_{\bar{v}}(\mu)
\end{equation}
defines a finite-time friction coefficient. In the quasistatic limit $\mu\to 0$, it reduces to the conventional Green–Kubo thermodynamic friction coefficient. For other protocols, Eq.~\eqref{eq:friction_coef}  can be interpreted as an average friction coefficient. In the quasistatic limit, we expect this coefficient to exhibit vastly different behavior in different regimes, according to Table~\ref{tab_chaos}. In regular and marginally unstable systems, the friction coefficient vanishes, while it saturates to a finite but nonzero value in chaotic, thermalizing systems. In contrast, this coefficient is expected to diverge in chaotic, nonthermalizing systems. Measurements of the dissipated work under slow cyclic driving, therefore, provide direct access to the low-frequency spectral response and, within this classification, distinguish regular from chaotic Hamiltonian dynamics. In certain cases, we found that it was easier to extract the Fisher information from the variance of $\Delta K_0$ instead of its mean. This is achieved through the generalized Jarzynski identity~\cite{jarzynski_1997}
\begin{equation}
    \left\langle e^{-\Delta K_0} \right\rangle = 1,
\end{equation}
which can be expanded in cumulants to get
\begin{equation}
    \text{Var}(\Delta K_0)\approx {\bar{v}^2} \mathscr{I}_{\bar{v}}(\mu) ,
    \label{eq_Evar}
\end{equation}
when other higher-order cumulants of $\Delta K_0$ are negligible.

In non-Hamiltonian autonomous systems, the Gibbs entropy of the system, $S$, itself need not be conserved, where
\begin{equation}
    S = -\int d\mathbf{x} \ \rho(\mathbf{x})\ln\rho(\mathbf{x}).
\end{equation}
Thus, a more general version of Eq.~(\ref{eq_modEnergyChange}) is
\begin{equation}
    \langle \Delta K_0 \rangle - \Delta S \simeq \frac{1}{2}\mathscr{I}_{\bar{v}}(\mu)\bar{v}^2.
\end{equation}
Volume-preserving dynamics guarantees $\Delta S=0$, whereas compressible flows generally produce an additional entropy contribution. Thus, the speed-Fisher information retains an interpretation in terms of the relative entropy generated by the protocol, although it does not generally admit an interpretation as dissipated work or thermodynamic friction. It may nevertheless be inferred from measurements of the final distribution whenever the corresponding relative entropy is experimentally accessible.

\section{Quantum to Classical Crossover}
\label{sec_crossover}

In a quantum system, the expression for the speed-Fisher information, given by Eq.~(\ref{eq_qsfi}), is valid only when the condition in Eq.~(\ref{eq_semiClassCond}) is satisfied. This condition can correspond to either the semiclassical limit $\hbar\to 0$, high-temperature limit, or the quasistatic limit $\mu \to 0$. When neither of these conditions is met, the quantum speed-Fisher information has new scaling regimes that are not possible in classical systems. In such a case, the scaling of the quantum speed-Fisher information with $\mu$ is manifestly a quantum effect.

To study this crossover, we consider a chaotic many-body system initially prepared in a Gibbs state at temperature $T$. The system is assumed to be sufficiently large so that its Heisenberg time is much longer than the timescales being probed. Therefore, the spectral function of any observable appears smooth on such timescales. The speed-Fisher information in Eq.~(\ref{eq_fisherInf}) can be exactly expressed as
\begin{equation}
    \mathscr{I}_{\bar{v}}(\mu) = \frac{4}{\hbar^2\mu^3} \int_{-\infty}^\infty \frac{d\omega}{2\pi} \tanh^2\left(\frac{\hbar\mu\omega}{2\kB T}\right)\tilde{C}_{V}(\mu \omega) \left|\tilde{f}\left(\omega\right)\right|^2.
\end{equation}
For concreteness, we consider a typical scenario, where the autocorrelation of the perturbation $V$ decays exponentially at long times with rate $\Gamma$ (see, for example, Refs.~\cite{Bulchandani_2022,kim_2025}). We model the spectral function of $V$ in the relevant frequency range by the Lorentzian form
\begin{equation}
    \tilde{C}_{V}(\omega) = \frac{2C_0}{\tau_{\text{rel}}}\frac{1}{(\omega^2 + \nicefrac{1}{\tau_{\text{rel}}^2})},
\end{equation}
where $\tau_{\text{rel}}=\nicefrac{1}{\Gamma}$ is the relaxation time. Additionally, if we consider an exponentially modulated protocol given by Eq.~(\ref{eq_expProtocol}), the speed-Fisher information becomes
\begin{equation}
\label{eq:FI_exact_exponential_decay}
    \mathscr{I}_{\bar{v}}(\mu) = \frac{8C_0}{\hbar^2\mu^3\tau_{\text{rel}}} \int_{-\infty}^\infty \frac{d\omega}{2\pi} \frac{\tanh^2\left(\frac{\hbar\mu\omega}{2\kB T}\right)}{(\mu^2\omega^2 + \nicefrac{1}{\tau_{\text{rel}}^2})(1 + \omega^2)^2}.
\end{equation}

The integral can be evaluated exactly, as shown in Appendix~\ref{app_quantumScaling}. Its asymptotic behavior is controlled by the interplay of three timescales: the protocol time $t_\mu=\nicefrac{1}{\mu}$, the relaxation time $\tau_{\text{rel}}$, and the Planckian time $\tau^{}_{\text{Pl}}=\nicefrac{\hbar}{\kB T}$~\cite{hartnoll_2022}. The leading expressions for the different timescale orderings are summarized in Table~\ref{tab_regimes}.

\begin{table}[htb]
\begin{ruledtabular}
\begin{tabular}{@{}r@{\hspace{0.4em}}ccc@{}}
    & & \multicolumn{2}{c}{$\displaystyle
      \frac{(\kB T)^2}{C_0}\mathscr{I}_{\bar v}(\mu)$}
    \\ \cmidrule(lr){3-4}
    & Timescale ordering & Prefactor & Time dependence
    \\ \midrule

    i)
    & $t_\mu\ll\tau_{\text{rel}}\ll\tau^{}_{\text{Pl}}$
    & $4$
    & $\displaystyle\frac{t_\mu^4}{\tau^{2}_{\text{Pl}}}$
    \\ \addlinespace[0.5em]

    ii)
    & $t_\mu\ll\tau^{}_{\text{Pl}}\ll\tau_{\text{rel}}$
    & $\displaystyle\frac{56\zeta(3)}{\pi^3}\approx 2.171$
    & $\displaystyle
       \frac{t_\mu^4}{\tau_{\text{rel}}\tau^{}_{\text{Pl}}}$
    \\
    \noalign{%
        \vskip0.5em
        \hbox to\linewidth{%
            \xleaders\hbox{\rule{3pt}{0.4pt}\kern2pt}\hfill
        }%
        \vskip0.5em
    }

    iii)
    & $\tau_{\text{rel}}\ll t_\mu\ll\tau^{}_{\text{Pl}}$
    & $2$
    & $\displaystyle
       \frac{\tau_{\text{rel}}}{\tau^{2}_{\text{Pl}}}t_\mu^3$
    \\
    \specialrule{0.9pt}{0.5em}{0.5em}

    iv)
    & $\tau^{}_{\text{Pl}}\ll t_\mu\ll\tau_{\text{rel}}$
    & $\displaystyle\frac{1}{2}$
    & $\displaystyle\frac{t_\mu^3}{\tau_{\text{rel}}}$
    \\
    \noalign{%
        \vskip0.5em
        \hbox to\linewidth{%
            \xleaders\hbox{\rule{3pt}{0.4pt}\kern2pt}\hfill
        }%
        \vskip0.5em
    }

    v)
    & $\tau_{\text{rel}}\ll\tau^{}_{\text{Pl}}\ll t_\mu$
    & $\displaystyle\frac{1}{2}$
    & $\tau_{\text{rel}}t_\mu$
    \\ \addlinespace[0.5em]

    vi)
    & $\tau^{}_{\text{Pl}}\ll\tau_{\text{rel}}\ll t_\mu$
    & $\displaystyle\frac{1}{2}$
    & $\tau_{\text{rel}}t_\mu$
    \\
\end{tabular}
\end{ruledtabular}
\caption{Leading asymptotic behavior of the speed-Fisher information for a Lorentzian spectral function and an
exponentially modulated protocol. The product of the last two columns gives the leading expression for
$(\kB T)^2\mathscr{I}_{\bar v}(\mu)/C_0$. The thick solid line separates regimes with
$t_\mu\ll\tau^{}_{\text{Pl}}$ above from those with $\tau^{}_{\text{Pl}}\ll t_\mu$ below. Only the $t_\mu^3$ and $t_\mu$ asymptotes below this line survive in the classical limit $\hbar\to0$ at fixed
temperature and protocol time, although they also apply to quantum systems. Dashed lines separate the $t_\mu^4$,
$t_\mu^3$, and $t_\mu$ scaling regimes. Here $t_\mu=\mu^{-1}$ and $\zeta(z)$ denotes the Riemann zeta function.}
\label{tab_regimes}
\end{table}

For protocol times shorter than both intrinsic timescales, the speed-Fisher information grows as $t_\mu^4$. When the relaxation and Planckian times are well separated, an intermediate $t_\mu^3$ regime emerges for protocol times that are in between the two intrinsic timescales. Although the exponent of $t_\mu^3$ is the same for either ordering of $\tau_{\text{rel}}$ and $\tau_{\text{Pl}}$, the prefactors are different, as shown in rows (iii) and (iv) of Table~\ref{tab_regimes}. For protocol times longer than both intrinsic timescales, the response grows linearly with $t_\mu$, recovering the quasistatic scaling. Thus, well-separated relaxation and Planckian times produce successive $t_\mu^4$, $t_\mu^3$, and $t_\mu$ regimes. When the two intrinsic timescales are comparable, there is no parametrically broad intermediate $t_\mu^3$ regime.

The classical speed-Fisher information for a Lorentzian spectral function follows from taking the limit $\hbar\to 0$ in Eq.~\eqref{eq:FI_exact_exponential_decay}:
\begin{equation}
    \mathscr{I}_{\bar{v}}(\mu) = \frac{2C_0}{\mu(\kB T)^2\tau_{\text{rel}}} \int_{-\infty}^\infty \frac{d\omega}{2\pi} \frac{\omega^2}{(\mu^2\omega^2 + \nicefrac{1}{\tau_{\text{rel}}^2})(1 + \omega^2)^2}.
\end{equation}
This integral can again be exactly solved to obtain
\begin{equation}
    \mathscr{I}_{\bar{v}}(\mu) = \frac{C_0\tau_{\text{rel}}}{2(\kB T)^2 \mu (\mu\tau_{\text{rel}} + 1)^2}.
\end{equation}
The two asymptotic regimes of this expression are given by rows (iv) and (vi) of Table~\ref{tab_regimes}. These are the only regimes that survive in the classical limit
$\hbar\to0$ at fixed temperature and protocol time. However, they also describe quantum systems when $\tau^{}_{\text{Pl}}\ll t_\mu$, with $C_0$ and $\tau_{\text{rel}}$ retaining their dependence on the quantum equilibrium state and dynamics. For a Lorentzian spectral function, the corresponding $t_\mu^3$ and $t_\mu$ scalings coincide with those of the exponentially
regularized AGP norm~\cite{tokarczyk2026}. In contrast, the $t_\mu^4$ regime requires $t_\mu\ll\tau^{}_{\text{Pl}}$ and disappears in the classical limit at fixed protocol time. In this regime, the speed-Fisher information depends differently on protocol time than the regularized AGP norm.

\begin{figure*}[htb]
    \centering
    \begin{subfigure}[t]{0.47\linewidth}
        \centering
        \includegraphics[]{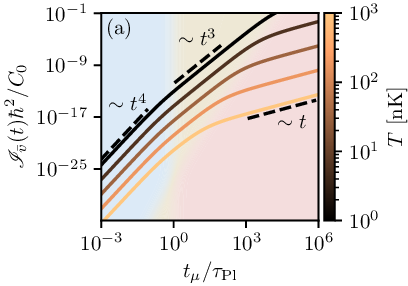}
        \phantomcaption{}
        \label{fig_qfi}
    \end{subfigure}
    \hfill
    \begin{subfigure}[t]{0.47\linewidth}
        \centering
        \includegraphics[]{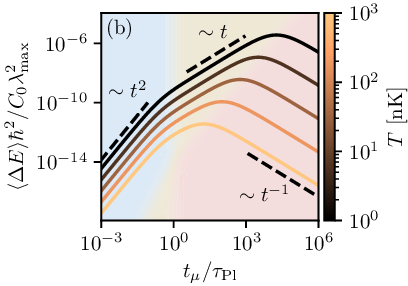}
        \phantomcaption{}
        \label{fig_dissE}
    \end{subfigure}
    \begin{subfigure}[t]{0.47\linewidth}
        \centering
        \includegraphics[]{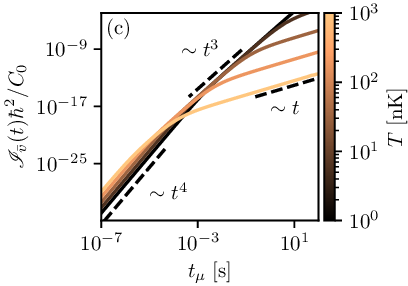}
        \phantomcaption{}
        \label{fig_qfi2}
    \end{subfigure}
    \hfill
    \begin{subfigure}[t]{0.47\linewidth}
        \centering
        \includegraphics[]{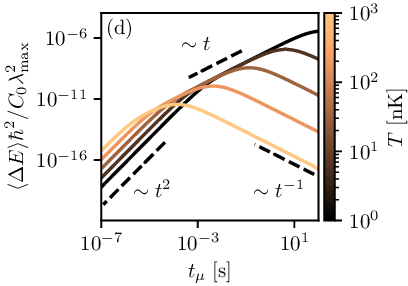}
        \phantomcaption{}
        \label{fig_dissE2}
    \end{subfigure}
    \caption{\raggedright (a) The speed-Fisher information as a function of protocol time $t_\mu$, in units of the Planckian time, for a cold Fermi gas with a Fermi temperature of $1$ $\mu$K and $k_F|a|=0.1$. The relaxation time is given by $\tau_{\text{rel}} = \frac{9\hbar T_F}{16\pi\kB(k_Fa)^2}\frac{1}{T^2}$ in the weak coupling limit of this system~\cite{bruun_2011}. We observe a transition from $\sim t_\mu^4$ to $\sim t_\mu^3$ and, finally, to $\sim t_\mu$ as the protocol time increases. The blue, yellow, and red regions correspond to fast, intermediate, and slow driving, respectively. (b) The corresponding scaling of the dissipated energy for a protocol with fixed amplitude $\lambda_{\text{max}}$. (c~\&~d) The same plots of speed-Fisher information and dissipated energy as a function of the unscaled time.}
\end{figure*}

The results summarized in Table~\ref{tab_regimes} reveal several crossovers in the speed-Fisher information. Its temperature dependence enters through the overall factor $C_0(T)/(\kB T)^2$, the Planckian time $\tau^{}_{\text{Pl}}$, and the system-dependent relaxation time $\tau_{\text{rel}}$. For a specific illustration, we use a Lorentzian model with the low-temperature relaxation law of a weakly interacting, three-dimensional Fermi gas in the normal phase. The restricted scattering phase space gives $\tau_{\text{rel}}\propto T^{-2}$~\cite{duine_2011,bruun_2011}. At low temperatures, the relevant hierarchy is therefore
$\tau_{\text{Fermi}}\ll\tau^{}_{\text{Pl}}\ll\tau_{\text{rel}}$, where $\tau_{\text{Fermi}}=\hbar/E_F$ is the microscopic timescale set by the Fermi energy.

The exact quantum speed-Fisher information for this Fermi gas model, given by Eq.~(\ref{eq_exactQFI}), is plotted at different temperatures in Figs.~\ref{fig_qfi} and~\ref{fig_qfi2}. Panel (a) shows the same data as panel (c), but as a function of the dimensionless protocol time $t_\mu/\tau^{}_{\text{Pl}}$. As the protocol time increases, the response crosses from regime (ii), with $\mathscr{I}_{\bar v}\sim t_\mu^4$, through regime (iv), with $\mathscr{I}_{\bar v}\sim t_\mu^3$, and finally to
regime (vi), with $\mathscr{I}_{\bar v}\sim t_\mu$. Since $\tau_{\text{rel}}/\tau^{}_{\text{Pl}}\propto T^{-1}$,
the intermediate regime becomes parametrically broader as the temperature decreases. At the same time, the
onset of the linear-in-$t_\mu$ quasistatic response is pushed to longer protocol times.

It is also instructive to examine the dissipated energy for fixed-amplitude protocols, shown in Figs.~\ref{fig_dissE} and~\ref{fig_dissE2}. For the exponentially modulated protocol,
$\bar v=2\lambda_{\rm max}/t_\mu$. In the regime $t_\mu\gg\tau^{}_{\text{Pl}}$, the relation between the dissipated energy and the speed-Fisher information gives
\begin{equation}
    \langle\Delta E\rangle
    \simeq
    \frac{C_0(T)\lambda_{\rm max}^2}{\kB T}
    \frac{\tau_{\text{rel}}t_\mu}
         {(t_\mu+\tau_{\text{rel}})^2}.
\end{equation}
Thus, the dissipated energy increases linearly with $t_\mu$ for $\tau^{}_{\text{Pl}}\ll t_\mu\ll\tau_{\text{rel}}$, but decreases as $t_\mu^{-1}$ for $t_\mu\gg\tau_{\text{rel}}$, recovering the adiabatic limit. To leading order in $\tau^{}_{\text{Pl}}/\tau_{\text{rel}}$, it reaches its maximum at
\begin{equation}
    t_{\mu,\max}\simeq\tau_{\text{rel}}\propto T^{-2},
    \qquad
    \langle\Delta E\rangle_{\max}
    \simeq
    \frac{C_0(T)\lambda_{\rm max}^2}{4\kB T}.
\end{equation}
For temperature-independent $C_0$, cooling therefore shifts the maximum to later times and increases its height as $T^{-1}$. More generally, the peak height scales as $C_0(T)/T$, so its temperature dependence cannot be inferred from the relaxation time alone.

The enhancement of sufficiently slow-driving dissipation reflects the increasing correlation time: the generalized friction coefficient scales as $\zeta_{\rm GK}=C_0(T)\tau_{\text{rel}}/(\kB T)$. A closely related mechanism underlies the increase of normal-state Fermi-liquid viscosity as the scattering rate decreases~\cite{bruun_2011}. This does not imply that colder systems dissipate more at every fixed protocol time, because cooling also moves the onset of the slow-driving regime to longer times.

For temperature-independent $C_0$, colder gases therefore dissipate more for sufficiently long protocols, provided $t_\mu\gg\tau_{\text{rel}}$ at all temperatures being compared. This enhancement does not persist indefinitely when the temperature is lowered at a fixed protocol time. Since $\tau_{\text{rel}}\propto T^{-2}$, sufficiently strong cooling eventually makes the relaxation time longer than the protocol duration, and further cooling suppresses dissipation. Thus, the enhanced long-time friction and the suppressed response at sufficiently low temperature are compatible: the onset of the slow-driving regime is pushed to increasingly long times as the gas is cooled.

Within the Lorentzian relaxation model considered here, the dissipative contribution vanishes as $T\to0$ at fixed $t_\mu$, provided the spectral weight $C_0(T)$ remains bounded. In this sense, the dynamics probed by a finite-duration protocol approach an effectively integrable limit, consistent with the correspondence
between decreasing temperature and approaching integrability, discussed in Ref.~\cite{kim2026temperatureintegrability}.

\section{Numerical Examples}
\label{sec_num}
We now test the relation between slow-driving response and low-frequency spectral weight in several classical and quantum systems. In all the examples below, we use the sinusoidal cyclic protocol given by Eq.~(\ref{eq_drive}), which has a duration of $t_\mu=\nicefrac{1}{\mu}$ and average speed of $\bar v$. For each $\mu$, we evaluate the change in modular energy at several small values of $\bar v$ and verify that the ratios entering Eqs.~\eqref{eq_modEnergyChange} and~\eqref{eq_Evar} become independent of $\bar v$. We then use these ratios to extract the $\bar v\to0$ speed-Fisher information. Repeating this procedure as $\mu$ is decreased determines the quasistatic scaling of $\mathscr{I}_{\bar v}(\mu)$. In all the examples below, we evolved the quantum states using a Krylov-subspace method~\cite{park_1986} and integrated the classical trajectories using a fourth-order symplectic Runge-Kutta integrator~\cite{calvo_1993}.

\begin{figure}[htb]
    \centering
    \includegraphics[]{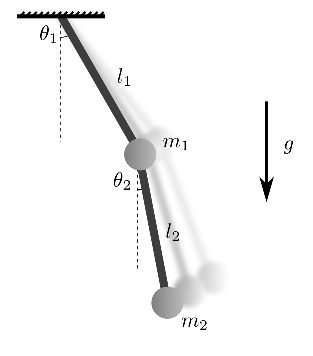}
    \caption{\raggedright Sketch of a double pendulum.}
    \label{fig_dbp}
\end{figure}

\begin{figure*}[htb]
    \centering
    \begin{subfigure}[t]{0.47\linewidth}
        \includegraphics[]{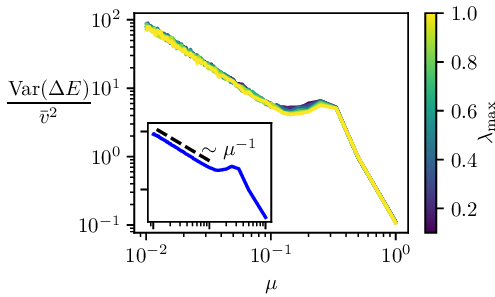}
        \caption{}
        \label{fig_dbp_fixL}
    \end{subfigure}
    \hfill
    \begin{subfigure}[t]{0.47\linewidth}
        \includegraphics[]{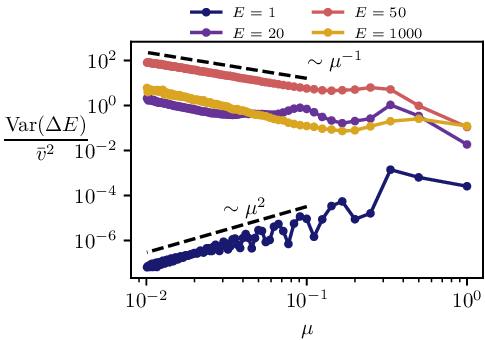}
        \caption{}
        \label{fig_dbp_E}
    \end{subfigure}
    \caption{\raggedright (a) $\nicefrac{\operatorname{Var}(\Delta E)}{\bar v^2}$ as a function of $\mu$ at several fixed protocol amplitudes $\lambda_{\text{max}}$ for the double pendulum system with energy $E=50$. The response is approximately independent of $\lambda_{\text{max}}$. Inset: $\nicefrac{\operatorname{Var}(\Delta E)}{\bar v^2}$ averaged over all $\lambda_{\text{max}}$s. (b) $\nicefrac{\operatorname{Var}(\Delta E)}{\bar v^2}$ at different energies of the double pendulum. The $E=20$ and $E=1000$ curves scale as $\sim\mu^{-1.5}$ and $\sim\mu^{-1.7}$ respectively. Parameters used: $m_1=1$, $m_2=1$, $l_1=1$, $l_2=1$, $g=9.81$, energy shell width $\sigma=0$.}
\end{figure*}

\begin{figure*}[htb]
    \centering
    \includegraphics[]{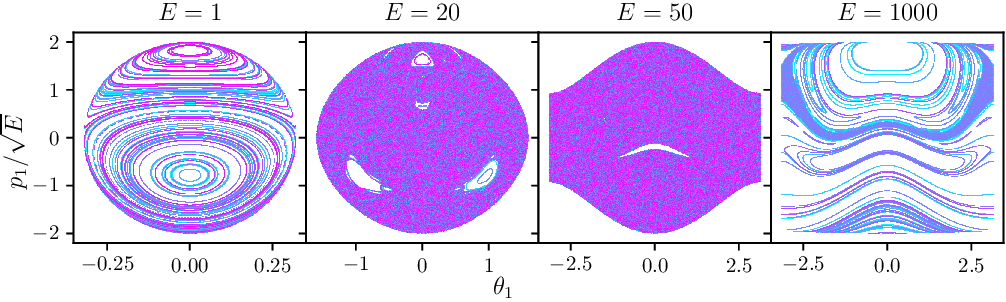}
    \caption{The $\theta_2=0$ Poincar\'e section of the double pendulum at energies $E=1$, $20$, $50$, and $1000$.}
    \label{fig_dbpPoincare}
\end{figure*}

\subsection{Classical Double Pendulum}

Our first example is the classical double pendulum, a paradigmatic nonlinear Hamiltonian system whose mixed phase space supports both regular and chaotic motion~\cite{shinbrot_1992}. Assuming $m_1$ and $m_2$ correspond to the two masses, $l_1$ and $l_2$ are the lengths of the rods, $\theta_1$ and $\theta_2$ are their angles from the vertical, and $p_1$ and $p_2$ are the corresponding canonical momenta (see Fig.~\ref{fig_dbp}), its Hamiltonian is given by~\cite{jimenez-lopez_2024}
\begin{align}
    H_0 =& \frac{1}{\nu}\left(\frac{p_1^2}{2Ml_1^2} + \frac{p_2^2}{2m_2l_2^2} - \frac{p_1p_2\cos\Delta\theta}{Ml_1l_2}\right)\notag\\& + Mgl_1(1-\cos\theta_1) + m_2gl_2(1-\cos\theta_2),
\end{align}
where $\Delta\theta = \theta_1-\theta_2$, $M = m_1+m_2$, and $\nu = 1  - \frac{m_2}{M}\cos^2\Delta\theta$. This system has two integrable limits: (i) When the energy $E\to 0$, the pendulum performs small oscillations about its minima, which can be approximated as harmonic motion, and (ii) As $E\to\infty$, gravity becomes negligible and the Hamiltonian depends on the angles only through $\theta_1-\theta_2$. The resulting conservation of $p_1+p_2$, together with energy conservation, makes this limiting system integrable.
At intermediate energies, the pendulum exhibits chaotic dynamics with a positive Lyapunov exponent. Thus, we expect the speed-Fisher information of the pendulum to exhibit different scalings at different energies.

We cyclically modulate the effective gravitational acceleration $g$, which can be implemented through a prescribed vertical acceleration of the pivot. The corresponding perturbation is
\begin{equation}
    V = Ml_1(1-\cos\theta_1) + m_2l_2(1-\cos\theta_2).
\end{equation}
The system is initialized in a regularized microcanonical state at energy $E$, with
\begin{equation}
    \rho(\theta,p) \propto e^{-\frac{1}{2\sigma^2}(H_0(\theta,p) - E)^2}.
\end{equation}
Here $\sigma$ is a small but finite energy width. The modular Hamiltonian is given by $K_0 = \frac{1}{2\sigma^2}(H_0 - E)^2 + \text{ const}$, and thus, according to Eq.~(\ref{eq_modEnergyChange}), in the asymptotic regime $\sigma\to 0$,
\begin{equation}
    \mathscr{I}_{\bar v}(\mu)
    \simeq\frac{1}{\sigma^2}\lim_{\bar v\to0}
    \frac{\operatorname{Var}(\Delta E)}{\bar v^2}.
\end{equation}

In Fig.~\ref{fig_dbp_E}, we plot $\nicefrac{\operatorname{Var}(\Delta E)}{\bar v^2}$ information of the double pendulum initialized at different energies. As expected, $\nicefrac{\operatorname{Var}(\Delta E)}{\bar v^2}$ vanishes as $\mu\to 0$ at a small energy reflecting regular motion of the pendulum. Moreover, in the linearized regime, the spectral function is composed of delta-function peaks, i.e.
\begin{equation}
    \tilde{\mathcal{D}}_V(\omega) = \sum_i A_i \delta(\omega - \omega_i).
\end{equation}
Plugging this into the expression for the speed-Fisher information of the sinusoidal protocol, we find that
\begin{equation}
    \mathscr{I}_{\bar v}(\mu)\simeq 8\pi^3\mu^2\sum_i \frac{A_i}{\omega_i^4}
    \sin^2\left(\frac{\omega_i}{2\mu}\right).
\end{equation}
Thus, at small $\mu$ the response has an envelope proportional to $\mu^2$, with oscillations determined by the spectral frequencies $\omega_i$, consistent with the observed low-energy behavior. This behavior is also consistent with the Poincar\'e section of the double pendulum being composed of regular trajectories at this energy (Fig.~\ref{fig_dbpPoincare}).

At a slightly higher energy of $E=20$, the speed-Fisher information exhibits a small initial dip, which is reminiscent of a regular motion at intermediate times (for a lower energy we expect that this initial dip will extend to lower values of $\mu$), while ultimately diverging at smaller values of $\mu$. This divergence is faster than $\nicefrac{1}{\mu}$, putting it in the nonthermalizing category. Such behavior is consistent with slow dynamics associated with a mixed phase-space -- as suggested by the corresponding Poincar\'e section (Fig.~\ref{fig_dbpPoincare}), which contains chaotic regions interspersed with islands of stability.

At an even higher energy of $E=50$, almost all of the islands of stability vanish, and the speed-Fisher information is observed to diverge linearly with $\sim t_\mu$ consistent with the asymptotic regime (vi) in Table~\ref{tab_regimes}. Pushing towards an even larger energy of $E=1000$ brings us closer to another integrable limit, and the Poincar\'e section again becomes a mix of chaotic and regular trajectories. Thus, the speed-Fisher information again shows signs of weak integrability breaking.

\subsection{Two-Spin Model}

\begin{figure*}[htpb]
    \centering
    \begin{subfigure}[t]{0.47\textwidth}
        \centering
        \includegraphics[]{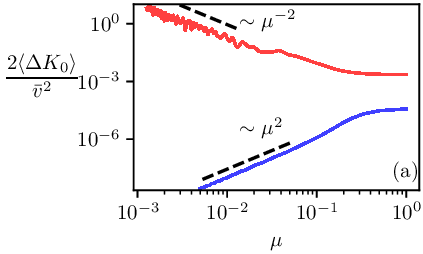}
        \phantomcaption
        \label{fig_work2Spin}
    \end{subfigure}
    \hfill
    \begin{subfigure}[t]{0.47\textwidth}
        \centering
        \includegraphics[]{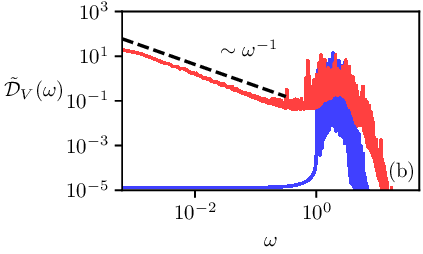}
        \phantomcaption
        \label{fig_specFn2Spin}
    \end{subfigure}
    \caption{\raggedright (a) The speed-Fisher information, $2\langle\Delta K_0\rangle/\bar{v}^2$, as a function of $\mu$ for the Gibbs ensemble, with $\kB T= 100$, of the integrable (blue) and chaotic (red) quantum two-spin models, with $S=10$. (b) The corresponding normalized classical spectral functions $\tilde{\mathcal{D}}_V(\omega)$. Parameters used: integrable model -- $\mathbf{J} = (1,1,1/2)$, $\mathbf{A} = (0,0,0)$, chaotic model -- $\mathbf{J} = (3/2,\pi,\sqrt{e})$, $\mathbf{A} = (\sqrt{\pi},\sqrt{3},e)$.}
\end{figure*}

We next turn to a minimal semiclassical setting in which the quantum response can be compared directly with the classical spectrum. This system consists of two spins, described by the Hamiltonian
\begin{align}
    H_0 = \frac{1}{S(S+1)}\sum_{\alpha \in\{x,y,z\}} \Big[&-J_{\alpha} S_{1\alpha} S_{2\alpha}\notag\\& + \frac{1}{2} A_\alpha \left(S_{1\alpha}^2 + S_{2\alpha}^2\right)\Big],
\end{align}
where $\mathbf{S}_1 = (S_{1x},S_{1y},S_{1z})$ and $\mathbf{S}_2 = (S_{2x},S_{2y},S_{2z})$ are spin-$S$ degrees of freedom in the quantum version, and are unit vectors in the classical case. Depending on the choice of parameters $\mathbf{J} = (J_x,J_y,J_z)$ and $\mathbf{A}=(A_x,A_y,A_z)$, this model can be integrable or chaotic~\cite{kim_2026,srivastava_1990}. In particular, the model is integrable if the parameters satisfy~\cite{srivastava_1990}
\begin{align}
    &(A_x-A_y)(A_y-A_z)(A_z-A_x) \notag\\& + \sum_{\alpha,\beta,\gamma = \text{cycl}(x,y,z)} J_\alpha^2(A_\beta-A_\gamma) = 0,
\end{align}
and is chaotic otherwise.

We prepare a Gibbs state and couple the drive to $V=S_{1z}S_{2z}$. Figure~\ref{fig_work2Spin} shows that the susceptibility decreases toward zero for the integrable model, whereas it grows approximately as $t_\mu^2$ for the chaotic one over the time window probed. The classical spectra in Fig.~\ref{fig_specFn2Spin} explain this contrast: the integrable spectrum falls to the numerical floor at low frequencies, while the chaotic spectrum exhibits approximate $1/|\omega|$ behavior, commonly called pink noise. These exponents are consistent with the scaling relation derived above and with the nonthermalizing behavior of the chaotic two-spin model.

The near-quadratic growth of the speed-Fisher information is the counterpart of the maximal growth of the regularized AGP norm discussed in Ref.~\cite{kim_2026}. According to Eq.~\eqref{eq:friction_coef}, it corresponds to, also maximal, approximately linear increase of the finite-time friction coefficient $\zeta_{\text{eff}}(\mu)$ with $t_\mu$. This example supports the predicted correspondence between the finite-$S$ quantum response and the classical low-frequency spectrum over the numerically accessible time window.

\subsection{Bose--Hubbard Model}

We now consider a system that provides a natural setting for experimentally probing the speed-Fisher information. Ultracold bosonic atoms confined in optical lattices~\cite{ketterle_2002,anglin_2002,bloch_2008,malvania_2021} are accurately described by the Bose--Hubbard model~\cite{gersch_1963}, whose hopping strength, interaction strength, and external confinement can all be controlled experimentally~\cite{gross_2017,jaksch_1998,jaksch_2005}. Energy absorption under lattice modulation has already been measured experimentally in Bose--Hubbard systems~\cite{rubio-abadal_2020}. The relations derived above show how measurements of dissipated energy under slow cyclic driving can be used to extract the speed-Fisher information. The Bose--Hubbard model, therefore, offers a realistic platform in which the connection between chaos, low-frequency spectral weight, and irreversible response may be tested.

\begin{figure*}[htb]
    \centering
    \begin{subfigure}[t]{0.47\textwidth}
        \centering
        \includegraphics[]{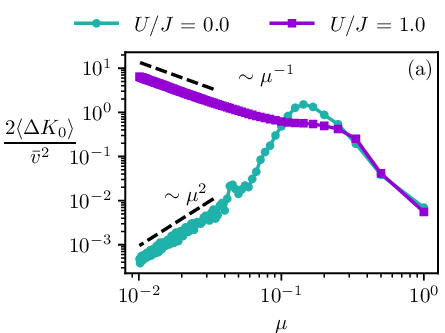}
        \phantomcaption
        \label{fig_hubbardFI}
    \end{subfigure}
    \hfill
    \begin{subfigure}[t]{0.47\textwidth}
        \centering
        \includegraphics[]{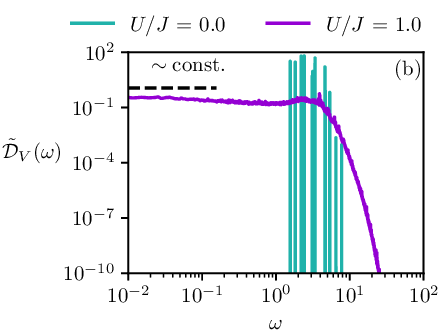}
        \phantomcaption
        \label{fig_hubbardSpec}
    \end{subfigure}
    \begin{subfigure}[t]{0.47\textwidth}
        \centering
        \includegraphics[]{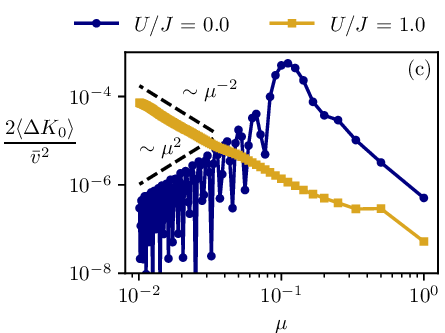}
        \phantomcaption
        \label{fig_hubbardFI3S}
    \end{subfigure}
    \hfill
    \begin{subfigure}[t]{0.47\textwidth}
        \centering
        \includegraphics[]{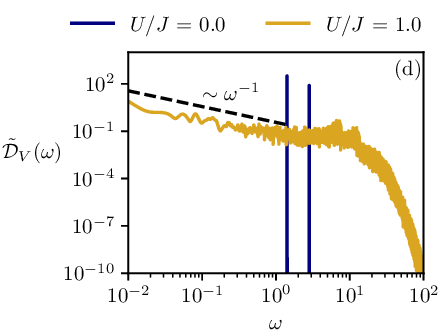}
        \phantomcaption
        \label{fig_hubbardSpec3S}
    \end{subfigure}
    \caption{\raggedright (a) The speed-Fisher information $\nicefrac{2\langle\Delta K_0\rangle}{\bar{v}^2}$ as a function of $\mu$ for the integrable ($U=0$, $J=1$) and chaotic ($U=1$, $J=1$) Bose--Hubbard model. (b) The corresponding spectral functions of the perturbation. Parameters used -- $N=L=8$, $k_0=1$, $\kB T =10$. (c~\&~d) The corresponding speed-Fisher information and spectral function plots for the 3-site Bose--Hubbard model. Parameters used -- $N=90$, $L=3$, $k_0=0$, $\kB T = 100$.}
\end{figure*}

We consider the following Bose--Hubbard Hamiltonian with $L$ sites and $N$ particles:
\begin{equation}
    H_0 = -J \sum_{i=0}^{L-1} b_i^\dagger b_{i+1} + \text{h.c.} + \frac{U}{2}\sum_{i=0}^{L-1} n_i(n_i-1) + \sum_{i=0}^{L-1} V_i n_i,
\end{equation}
with periodic boundary conditions. Here, $b_i^\dagger$ and $b_i$ represent the bosonic creation and annihilation operators at site $i$, $n_i=b_i^\dagger b_i$, $J$ describes the hopping strength, $U$ is the on-site repulsion term, and $V_i$ is a trapping potential. We assume this trapping potential to be harmonic, so that $V_i = \frac{1}{2}k_0\left(i-\frac{L-1}{2}\right)^2$. This model has two integrable limits: (i) the limit $\nicefrac{U}{J} \to 0$, which describes free, non-interacting bosons on the lattice, and (ii) the hard-core limit $\nicefrac{U}{J} \to \infty$, where the model maps to free fermions. For all other intermediate values of $\nicefrac{U}{J}$, the model is nonintegrable.

In this example, we choose $n=L=8$ and consider a periodically modulated external perturbation of the form
\begin{equation}
    V = \sum_{i=0}^{L-1} \cos\left[\frac{6\pi\left(i+\frac{1}{2}\right)}{L}\right] n_i.
\end{equation}
For $\nicefrac{U}{J}=0$, the Fisher information vanishes as $\sim\mu^2$ (Fig.~\ref{fig_hubbardFI}), reflecting the isolated peaks in the corresponding spectral function (Fig.~\ref{fig_hubbardSpec}). At $\nicefrac{U}{J}=1$, the spectral function exhibits a low-frequency plateau, producing the thermalizing scaling $\mathscr{I}_{\bar v}(\mu)\sim t_\mu$. This behavior is consistent with the GOE level statistics in the homogeneous Bose--Hubbard model
at comparable interaction strengths~\cite{kolovsky_2004,kollath_2010}.

In contrast to the thermalizing behavior of the eight-site example, the Bose--Hubbard trimer can exhibit mixed regular and chaotic dynamics. Its level-spacing statistics can lie between the Poisson and Wigner--Dyson distributions, reflecting the mixed phase space of its classical counterpart~\cite{nakerst_2023,bychek_2020}. We study this system in Figs.~\ref{fig_hubbardFI3S} and~\ref{fig_hubbardSpec3S}, using the perturbation $V=\nicefrac{n_0}{N}$.

We find that the speed-Fisher information decreases as $\mu^2$ in the noninteracting limit, whereas it grows approximately as $t_\mu^2 = \nicefrac{1}{\mu^2}$ in the interacting case over the time window studied. Through the spectral relation derived above, this near-quadratic growth is consistent with approximate $1/|\omega|$ pink-noise behavior at low frequencies. These results are consistent with the chaotic, nonthermalizing regime of our classification. Together, the Bose--Hubbard examples connect the spectral classification to a protocol that can be implemented in cold-atom experiments.

\subsection{One-Dimensional Mixed-Field Ising Model}

\begin{figure*}[htb]
    \centering
    \begin{subfigure}[t]{0.47\textwidth}
        \centering
        \includegraphics[]{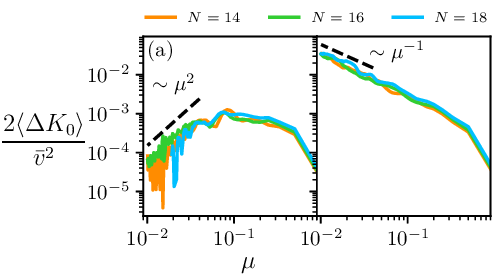}
        \phantomcaption
        \label{fig_work}
    \end{subfigure}
    \hfill
    \begin{subfigure}[t]{0.47\textwidth}
        \centering
        \includegraphics[]{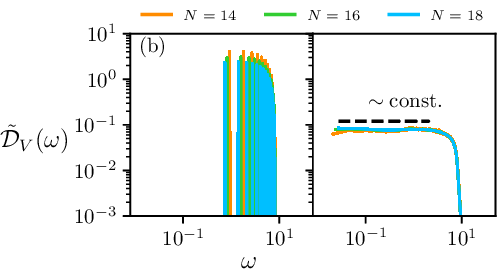}
        \phantomcaption
        \label{fig_specFn}
    \end{subfigure}
    \caption{\raggedright (a) The speed-Fisher information, $\nicefrac{2\langle\Delta K_0\rangle}{\bar{v}^2}$, as a function of $\mu$ for the Gibbs ensemble, with $\kB T= 100$, of the integrable (left) and chaotic (right) Ising models, with system sizes of $N=14$~(orange), $16$~(green), and $18$~(cyan). (b) The corresponding normalized spectral functions $\tilde{\mathcal{D}}_V(\omega)$. Parameters used -- integrable model: $J = 1$, $h_x=1.05$, $h_z=0$, chaotic model: $J = 1$, $h_x=1.05$, $h_z=0.5$.}
\end{figure*}

Another experimentally viable system to test the speed-Fisher information framework is the Ising model. Variants of this model can be realized using Rydberg-atom arrays~\cite{labuhn_2016}, trapped ions~\cite{kim_2011}, and superconducting-qubit systems~\cite{gong_2016}, in which the interaction strengths and applied fields are controllable in time. Here, we consider the one-dimensional, spin-$1/2$ Ising model with both transverse and longitudinal fields~\cite{ovchinnikov_2003}, described by the Hamiltonian:
\begin{equation}
    H = J \sum_{i=1}^N \sigma^z_i \sigma^z_{i+1} + h_x \sum_{i=1}^N \sigma^x_i + h_z \sum_{i=1}^N \sigma^z_i,
\end{equation}
with periodic boundary conditions ($\mathbf{\sigma}_1 = \mathbf{\sigma}_{N+1}$). In the absence of the longitudinal field ($h_z=0$), this model reduces to the transverse-field Ising chain, which is integrable~\cite{pfeuty_1970}. A longitudinal field generically breaks integrability, and the resulting mixed-field Ising model exhibits chaotic behavior~\cite{banuls_2011}. In either case, we drive this system through the perturbation $V = \displaystyle\sum_{i=1}^N \sigma^x_i$.

The two systems exhibit sharply different behavior in the slow-driving limit (Fig.~\ref{fig_work}). In the integrable case, the Fisher information decreases toward zero as $\mu \to 0$, whereas, in the chaotic thermalizing (ETH) regime, the Fisher information diverges as $\sim t_\mu=\nicefrac{1}{\mu}$. This behavior is consistent with the spectral function of $V$, plotted in Fig. \ref{fig_specFn}. While the spectral function of the integrable model exhibits discrete peaks, the chaotic case exhibits a low-frequency plateau. This suggests that the mixed-field Ising model belongs to the chaotic, thermalizing category.

\subsection{\texorpdfstring{$\beta$}{beta}--Fermi--Pasta--Ulam--Tsingou Model}

\begin{figure*}[htb]
    \centering
    \begin{subfigure}[t]{0.47\linewidth}
        \centering
        \includegraphics[]{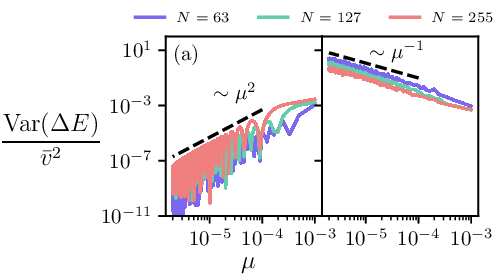}
        \phantomcaption
        \label{fig_betaDriven_a}
    \end{subfigure}
    \hfill
    \begin{subfigure}[t]{0.47\linewidth}
        \centering
        \includegraphics[]{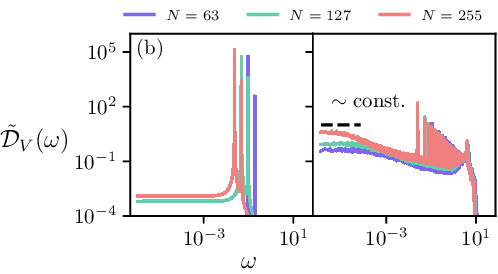}
        \phantomcaption
        \label{fig_betaDriven_b}
    \end{subfigure}
    \caption{\raggedright (a) $\nicefrac{\text{Var}(\Delta E)}{\bar{v}^2}$ as a function of $\mu$ in a classical $\beta$-FPUT model with $\beta=0$ (left) and $\beta=10$ (right), initialized in a microcanonical ensemble with $E=1$ and width $\sigma=0$. System sizes of $N=63$ (purple), $127$ (green), and $255$ (pink) are considered. (b) The corresponding spectral functions of the perturbation.}
    \label{fig_betaDriven}
\end{figure*}

We finally consider the $\beta$-Fermi--Pasta--Ulam--Tsingou ($\beta$-FPUT) model~\cite{fermi}, a classical many-body system whose study played a foundational role in understanding ergodicity and thermalization~\cite{campbell_2005,berman_2005,ford_1992}. Originally introduced to investigate how nonlinear interactions lead to energy equipartition, the FPUT model famously exhibits long-lived metastable behavior~\cite{benettin_2011,reiss_2023} and recurrences~\cite{tuck_1972,pierangeli_2018,pace_2019,karve_2024} in the weakly nonlinear regime, while sufficiently strong nonlinearities produce chaotic dynamics and rapid thermalization~\cite{onorato_2023,lvov_2015}. It therefore provides a natural setting in which to study how the irreversible response to slow driving changes between integrable and thermalizing regimes. This model describes a one-dimensional chain of oscillators through the Hamiltonian:
\begin{equation}
    H_0 = \sum_{n=1}^{N} \frac{p_n^2}{2} + \frac{1}{2}\sum_{n=0}^N (q_{n+1}-q_n)^2 + \frac{\beta}{4}\sum_{n=0}^N (q_{n+1}-q_n)^4,
\end{equation}
subject to fixed boundary conditions, $q_0=q_{N+1}=0$. For $\beta=0$, the model reduces to a harmonic chain and is integrable. For nonzero $\beta$, the quartic interaction breaks integrability. This model is known to exhibit metastability at small $\beta$, but thermalizes quickly at larger nonlinearities~\cite{benettin_2008,reiss_2023}. Here, we consider models with large numbers of particles, and apply a quartic perturbation 
\begin{equation}
\displaystyle\frac{\lambda(\tau)}{4} \sum_{n=0}^N (q_{n+1}-q_n)^4,    
\end{equation}
where $\lambda(\tau)$ is given by Eq.~(\ref{eq_drive}). We initialize this system in a microcanonical ensemble with energy $E=1$, which is, due to the large number of particles considered here, locally equivalent to a Gibbs state with temperature $\kB T = \left(\frac{\partial S}{\partial E}\right)^{-1}$~\cite{touchette_2015}. Then, Eq.~(\ref{eq_Evar}) allows us to extract the scaling of the speed-Fisher information with $\mu$ from $\nicefrac{\text{Var}(\Delta E)}{\bar{v}^2}$.

In Fig. \ref{fig_betaDriven}, we consider two extremes of the $\beta$-FPUT model: the integrable case with $\beta=0$, and a strongly nonlinear case with $\beta=10$. As expected, the thermodynamic drag vanishes in the integrable case as $\sim\mu^{2}$, and diverges as $\nicefrac{1}{\mu}$ when $\beta=10$. This is consistent with the behavior of the corresponding spectral function at small frequencies: in the integrable case, the spectral weight drops down to the numerical noise floor at small frequencies, whereas it approaches a finite low-frequency plateau in the nonintegrable case. Thus, the FPUT model with $\beta=10$ also belongs to the chaotic and thermalizing category.

\section{Conclusions}

In summary, in this article we discuss an operational notion of chaos that treats classical and quantum systems on an equal footing. We argue that chaos is linked to the response of a stationary state under slow driving, which, for specific protocols, is closely related to the regularized Fubini-Study geometric tensor. We introduce the speed-Fisher information, which quantifies the susceptibility of the stationary state to the average speed of the drive, and show that it can be interpreted as a thermodynamic drag. This Fisher information is also directly related to entropy generation and energy absorption. Additionally, we identify a quantum regime at times shorter than the Planckian timescale, in which the speed-Fisher information exhibits a uniquely quantum scaling behavior. The robustness of this framework is demonstrated by probing chaos via this drag in a variety of quantum and classical, few- and many-body systems.

In contrast to traditional perspectives, our framework posits that regular or chaotic behavior is a property not only of the system and its state, but also of the relevant observable, as reflected in its low-frequency spectral behavior. In chaotic thermalizing systems, most local observables are expected to reach equilibrium at the same time scale, known as the Thouless time~\cite{schiulaz_2019,suntajs_2020,suntajs_2022}. However, certain observables, such as currents, can relax faster, leading to a more stable response to perturbations coupled to them. Additionally, conserved perturbations (i.e. $[H_0,V]=0$) may completely fail to reveal the chaotic nature of the system, since they leave the stationary state unchanged. The difference between different observables becomes more pronounced close to integrable regimes~\cite{kim_2025}, as is often observed in real dynamical systems~\cite{mogavero_2023}. Our framework is therefore complementary to traditional diagnostics, offering not only a criterion for the presence of chaos, but also a classification into different universal classes and a detailed description of how chaos manifests in the response of different physical observables and timescales.

\textit{\textbf{Acknowledgments}} -- The authors thank Hyeongjin Kim, Guilherme Delfino, and Bernardo Barrera for insightful discussions. AP was supported by: NSF grant no. DMR-2412542 and AFOSR grant no. FA9550-21-1-0342. The authors acknowledge the use of Boston University’s Shared Computing Cluster (SCC) for numerical simulations.

\textit{\textbf{Data Availability Statement}} -- The data that support the findings of this study were generated by numerical simulations. The data, source code, and parameters used to generate the simulations are publicly available~\cite{karve_ising,karve_2026_beta_fput,karve_2026_dataset_2spin,karve_bh,karve_dbp}.

\appendix

\section{Quantum Speed-Fisher Information}
\label{app_qsfi}

In this section, we describe the details of the derivation of Eq.~(\ref{eq_qsfi}). We start from the definition of the quantum fidelity between the initial and final states:
\begin{equation}
    \mathscr{F}(\bar{v},\mu) = \left(\tr\sqrt{\rho_{-\infty}^2 + \sqrt{\rho_{-\infty}}\ \delta\rho \ \sqrt{\rho_{-\infty}}}\right)^2,
\end{equation}
where $\delta\rho = \rho_{+\infty}-\rho_{-\infty}$. The final state of the system $\rho_{+\infty}$ is given by the Dyson series:
\begin{align}
    \rho_{+\infty} =& \ \rho_{-\infty} + \frac{\bar{v}}{i\hbar\mu} \int_{-\infty}^{+\infty} d\tau \ f(\mu\tau) [V(\tau), \rho_{-\infty}] \notag\\& + \mathcal{O}\left(\bar{v}^2\right),
\end{align}
where $V(\tau) = e^{iH_0\tau/\hbar}Ve^{-iH_0\tau/\hbar}$ denotes Heisenberg evolution under $H_0$. To compute this fidelity, we define $X^2 = \rho_{-\infty}^2 + \sqrt{\rho_{-\infty}}\ \delta\rho \ \sqrt{\rho_{-\infty}}$, so that $\mathscr{F} = \left(\tr X\right)^2$. We express $X$ perturbatively in powers of the perturbation as
\begin{equation}
    X = X_0 + X_1 + X_2 + \mathcal{O}\left(\bar{v}^3\right).
\end{equation}
And thus, one can write
\begin{align}
    & X_0^2 + X_0 X_1 + X_1 X_0 + X_1^2 + X_0 X_2 + X_2 X_0 \notag\\& = \rho_{-\infty}^2 + \sqrt{\rho_{-\infty}}\ \delta\rho \ \sqrt{\rho_{-\infty}}.
\end{align}
Collecting all terms of the same order together, we get
\begin{subequations}
    \begin{equation}
        X_0 = \rho_{-\infty},
    \end{equation}
    \begin{equation}
        \rho_{-\infty} X_1 + X_1 \rho_{-\infty} = \sqrt{\rho_{-\infty}} \ \delta\rho \ \sqrt{\rho_{-\infty}},
    \end{equation}
    \begin{equation}
        X_1^2 + \rho_{-\infty}X_2 + X_2 \rho_{-\infty} = 0.
    \end{equation}
\end{subequations}
The zeroth-order identity $\tr X_0 = 1$. Plugging in $X_0=\rho_{-\infty}$ in the identity at first order, we get
\begin{align}
    &\rho_{-\infty} X_1 + X_1\rho_{-\infty} \notag\\& = \frac{\bar{v}}{2\mu} \int_{-\infty}^\infty d\tau \ f(\mu\tau)\left(\sqrt{\rho_{-\infty}} \ L_V(\tau) \rho_{-\infty} \ \sqrt{\rho_{-\infty}} \right. \notag\\&\qquad\qquad\qquad\qquad \left. + \sqrt{\rho_{-\infty}} \ \rho_{-\infty} L_V(\tau) \ \sqrt{\rho_{-\infty}} \right),
\end{align}
which then implies
\begin{equation}
    X_1 = \frac{\bar{v}}{2\mu} \int_{-\infty}^\infty d\tau \ f(\mu\tau) \sqrt{\rho_{-\infty}} \ L_V(\tau) \ \sqrt{\rho_{-\infty}}.
\end{equation}
It is easy to show that $\tr X_1 = 0$, using the cyclic property of the trace. Finally, the second order identity can be written as
\begin{equation}
    X_2 + \rho_{-\infty}^{-1} X_2 \rho_{-\infty} = -\rho_{-\infty}^{-1} X_1^2,
\end{equation}
assuming that $\rho_{-\infty}$ is invertible. Taking the trace of both sides gives us
\begin{align}
    \tr X_2 &= -\frac{1}{2} \tr(\rho_{-\infty}^{-1}X_1^2) \notag\\&= -\frac{\bar{v}^2}{8\mu^2} \int_{-\infty}^\infty d\tau_1 \int_{-\infty}^\infty d\tau_2 \ f(\mu\tau_1) f(\mu\tau_2) \notag\\&\qquad\qquad\qquad\qquad\qquad \tr(\rho_{-\infty} L_V(\tau_1)L_V(\tau_2)).
\end{align}
Note that the above expression is a double time-integral over the autocorrelation function of $L_V$. The fidelity can now be computed by combining all orders together:
\begin{align}
    \mathscr{F}(\bar{v},\mu) = 1 - \frac{\bar{v}^2}{4\mu^2} \int_{-\infty}^\infty d\tau_1\int_{-\infty}^\infty d\tau_2 \ & f(\mu\tau_1)f(\mu\tau_2) \notag\\& C_{L_V} (\tau_1-\tau_2).
\end{align}
In the Fourier space, this expression becomes
\begin{equation}
    \mathscr{F}(\bar{v},\mu) = 1 - \frac{\bar{v}^2}{4\mu^4} \int_{-\infty}^\infty \frac{d\omega}{2\pi} \ \tilde{C}_{L_V}(\omega) \left|\tilde{f}\left(\frac{\omega}{\mu}\right)\right|^2,
\end{equation}
from which, the speed-Fisher information in Eq.~(\ref{eq_fisherInf}) can be extracted.

To recast this expression in terms of the physical perturbation $V$, we write the spectral function of $L_V$ in the eigenbasis of $H_0$:
\begin{align}
    \tilde{C}_{L_V}(\omega) = \frac{4\pi}{\hbar^2}\sum_{m\neq n} \frac{[P(E_m)-P(E_n)]^2}{P(E_m)+P(E_n)} & \notag\\ |V_{mn}|^2 \delta(\omega-\omega_{mn}) &,
    \label{eq_clv}
\end{align}
where $\{E_n\}$ are the energy levels of $H_0$, and $\omega_{mn} = \frac{E_m-E_n}{\hbar}$. According to Eq.~(\ref{eq_fisherInf}), any singular dependence of the speed-Fisher information on $\mu$ in the limit $\mu\to 0$ is controlled by the low-frequency behavior of $\tilde{C}_{L_V}(\omega)$. Thus, if $P$ is a smooth function of energy, the low-frequency contribution, arising from transitions with $\omega_{mn}\sim\mu$, can be evaluated using
\begin{equation}
\frac{P(E_m)-P(E_n)}{P(E_m)+P(E_n)}
\approx
\frac{\hbar\omega_{mn}}{2}
\partial_E\ln P(\bar E_{mn}),
\end{equation}
where $\bar E_{mn}=(E_m+E_n)/2$. This approximation is valid only when Eq.~(\ref{eq_semiClassCond}) is satisfied. Then, the spectral function $\tilde{C}_{L_V}(\omega)$ can be approximated by $\omega^2 \tilde{\mathcal{D}}_V(\omega)$, where
\begin{align}
    \tilde{\mathcal{D}}_V(\omega) = \pi\sum_{m\neq n} (P(E_m)+P(E_n)) \left(\partial_E\ln P(\bar E_{mn})\right)^2 &\notag \\|V_{mn}|^2 \delta(\omega-\omega_{mn}),&
\end{align}
which we refer to as the score-weighted spectral function of $V$. When the condition in Eq.~\eqref{eq_semiClassCond} is satisfied, the score-weighted spectral function can, to leading order at low frequencies, be expressed as the Fourier transform of the score-weighted correlation function given in Eq.~(\ref{eq_DVt}). And therefore, we can write the speed-Fisher information in terms of $\tilde{\mathcal{D}}_V$ as
\begin{equation}
    \mathscr{I}_{\bar{v}}(\mu) \simeq \frac{1}{\mu}\int_{-\infty}^\infty \frac{d\omega}{2\pi} \ \tilde{\mathcal{D}}_V(\mu \omega) \left|\tilde{g}(\omega)\right|^2.
\end{equation}

\section{Classical Speed-Fisher Information}
\label{app_csfi}

In classical Hamiltonian systems, we start with the equation of motion:
\begin{equation}
    \frac{\partial\rho_{\tau}(\mathbf{x})}{\partial \tau} = \{H(\mathbf{x},\tau),\rho_{\tau}(\mathbf{x})\}.
\end{equation}
We assume that up to leading order in $\bar{v}$, this new distribution can be written as $\rho_{\tau}(\mathbf{x}) = \rho_{-\infty}(\mathbf{x})\left(1 + \frac{\bar{v}}{\mu} u_\tau(\mathbf{x})\right)$. Then, $u$ satisfies
\begin{equation}
    \frac{du_\tau(\mathbf{x})}{d\tau} - \{H_0(x),u_\tau(\mathbf{x})\} = f(\mu \tau) \{V(\mathbf{x}),\ln\rho_{-\infty}(\mathbf{x})\}.
\end{equation}
Since the left-hand side of the above equation is the derivative of $u$ along the unperturbed trajectory, we can write the solution as
\begin{equation}
    u_{+\infty}(\mathbf{x}) = \int_{-\infty}^{+\infty} d\tau \ f(\mu\tau) L_V(\tau).
\end{equation}
And therefore, the correction to the distribution at $t=\infty$ is given by
\begin{equation}
    \delta\rho(\mathbf{x}) = \frac{\bar{v}}{\mu} \rho_{-\infty}(\mathbf{x}) \int_{-\infty}^\infty d\tau \ f(\mu\tau) L_V(\tau).
\end{equation}
Thus, the classical fidelity between the states at $t=\pm\infty$ given by Eq.~(\ref{eq_classicalFid}) can be written in terms of $\delta\rho$:
\begin{equation}
    \mathscr{F}(\bar{v},\mu) = 1 - \frac{1}{4}\int d\mathbf{x} \ \frac{\delta\rho(\mathbf{x})^2}{\rho_{-\infty}(\mathbf{x})}.
\end{equation}
Consequently, the fidelity depends on the autocorrelation function of $L_V$:
\begin{align}
    \mathscr{F}(\bar{v},\mu) = 1 - \frac{\bar{v}^2}{4\mu^2}\int_{-\infty}^\infty d\tau_1 \int_{-\infty}^\infty d\tau_2 & \ f(\mu\tau_1)f(\mu\tau_2) \notag\\& C_{L_V}(\tau_1 - \tau_2).
\end{align}
And just like in the quantum case, the Fisher information can be written in terms of the spectral function of $L_V$ and the Fourier transform of $f$:
\begin{equation}
    \mathscr{I}_{\bar{v}}(\mu) = \frac{1}{\mu^3} \int_{-\infty}^\infty \frac{d\omega}{2\pi} \ \tilde{C}_{L_V}(\mu \omega) \left|\tilde{f}\left(\omega\right)\right|^2.
\end{equation}

\section{Speed-Fisher Information in Classical Autonomous Systems}
\label{app_nhsfi}

A classical autonomous system is described by an equation of motion of the form
\begin{equation}
    \frac{d\mathbf{x}}{d\tau} = \mathbf{F}(\mathbf{x}).
\end{equation}
We introduce a perturbation, so that the differential equation takes the form $\frac{d\mathbf{x}}{d\tau} = \mathbf{F}(\mathbf{x}) + \lambda(\tau)\mathbf{V}(\mathbf{x})$ and parameterize $\lambda$ as
\begin{equation}
   \lambda(\tau) =  \bar{v} f(\mu \tau), \text{ with } \int_{-\infty}^\infty |f(z)| \ dz = 1.
\end{equation}

As before, we initialize the system in $\rho_{-\infty}(\mathbf{x})$, which is a stationary probability distribution of $\mathbf{F}$, and let the system evolve over time. The new distribution satisfies the continuity equation:
\begin{equation}
    \frac{\partial \rho_\tau(\mathbf{x})}{\partial \tau} + \nabla\cdot [\rho_\tau(\mathbf{x})(\mathbf{F}(\mathbf{x}) + \bar{v} f(\mu \tau) \mathbf{V}(\mathbf{x}))] = 0.
\end{equation}
Like the classical Hamiltonian case, the difference between the states at $t=\pm\infty$, up to first order in $\bar{v}$, can be shown to be
\begin{equation}
    \delta\rho(\mathbf{x}) = - \bar{v} \rho_{-\infty}(\mathbf{x}) \int_{-\infty}^\infty d\tau \ f(\mu\tau) L_V(\mathbf{x}(\tau)), 
\end{equation}
where the logarithmic derivative is given by Eq.~(\ref{eq_logDerNonHam}). The speed-Fisher information can then be expressed in terms of $\tilde{C}_{L_V}$ as
\begin{equation}
    \mathscr{I}_{\bar{v}}(\mu) = \frac{1}{\mu}\int_{-\infty}^\infty \frac{d\omega}{2\pi} \ \tilde{C}_{L_V}(\mu \omega)|\tilde{f}(\omega)|^2.
\end{equation}

\section{Quasistatic Scaling of the Speed-Fisher Information}

\subsection{Regular Systems}

We assume that the score-weighted spectral function in regular systems is given by the ansatz in Eq.~(\ref{eq_specFnCond}) with $\alpha > 1$, and that $\tilde{g}(\omega)$ decays sufficiently fast, as given in Eq.~(\ref{eq_gDecay}).
Then, we split the integral in Eq.~(\ref{eq_qsfi}) as
\begin{align}
    \mathscr{I}_{\bar{v}}(\mu) =& \ \frac{1}{\mu} \int_{|\mu\omega| < \delta} \frac{d\omega}{2\pi} \ \tilde{\mathcal{D}}_{V}(\mu \omega) \left|\tilde{g}\left(\omega\right)\right|^2\notag\\& + \frac{1}{\mu} \int_{|\mu\omega| > \delta} \frac{d\omega}{2\pi} \ \tilde{\mathcal{D}}_{V}(\mu \omega) \left|\tilde{g}\left(\omega\right)\right|^2.
    \label{eq_i1i2}
\end{align}
Let us refer to the two terms in the above expression as $I_1$ and $I_2$, respectively. To show that $\mathscr{I}_{\bar{v}}(\mu)$ is finite in the limit $\mu\to 0$, it is sufficient to prove that both $I_1$ and $I_2$ are finite. We first find that $I_1$ satisfies
\begin{align}
    |I_1| &\leq \frac{1}{\mu} \int_{|\mu\omega| < \delta} \frac{d\omega}{2\pi} \ |\tilde{\mathcal{D}}_{V}(\mu \omega)| \left|\tilde{g}\left(\omega\right)\right|^2 \ \notag\\&\leq\  C_1C_2^2 \mu^{\alpha-1} \int_{|\mu\omega| < \delta} \frac{d\omega}{2\pi} \ |\omega|^{\alpha-2} \notag\\&= \frac{C_1C_2^2\delta^{\alpha-1}}{\pi(\alpha-1)}.
\end{align}
Thus $I_1$ is finite. Similarly, for $I_2$, we can write
\begin{align}
    |I_2| &\leq \frac{1}{\mu} \int_{|\mu\omega| > \delta} \frac{d\omega}{2\pi} \ |\tilde{\mathcal{D}}_{V}(\mu \omega)| \left|\tilde{g}\left(\omega\right)\right|^2 \notag\\& = \frac{1}{\mu^2} \int_{|\omega| > \delta} \frac{d\omega}{2\pi} \ |\tilde{\mathcal{D}}_{V}(\omega)| \left|\tilde{g}\left(\frac{\omega}{\mu}\right)\right|^2 \notag\\&\leq \frac{C_2^2}{\pi} \int_{\delta}^\infty d\omega \ \frac{|\tilde{\mathcal{D}}_{V}(\omega)|}{\omega^2}. 
\end{align}
Note that the integral $\int_{\delta}^\infty d\omega \ \frac{|\tilde{\mathcal{D}}_{V}(\omega)|}{\omega^2}$ has to be finite, since $\tilde{\mathcal{D}}_V(\omega\to\infty)$ cannot diverge. Thus, $I_2$ must also be finite.

\subsection{Chaotic and Marginally Unstable Systems}

We now repeat the above analysis for chaotic systems, where observables typically have nonzero low-frequency spectral weights~\cite{kim_2026}. The spectral function is again assumed to have the ansatz given by Eq.~(\ref{eq_specFnCond}), but with $\alpha \leq 0$. In thermalizing systems $\alpha=0$, and the spectral function plateaus at low frequencies. On the other hand, $\alpha < 0$ in nonthermalizing systems and the spectral function diverges as $\omega\to 0$. Moreover, there exists a lower bound on the allowed values of $\alpha$. To see this, consider the following integral:
\begin{equation}
    \int_{-\infty}^\infty \frac{d\omega}{2\pi} \tilde{\mathcal{D}}_V(\omega) = \left\langle s(H_0)^2 \delta V^2 \right\rangle.
\end{equation}
This integral must be finite, since the score-weighted variance of the observable $V$ must be finite. This can only be true when $-1 < \alpha$.

We can now split the speed-Fisher information into two separate integrals, as was done in Eq.~(\ref{eq_i1i2}). Then, we find that
\begin{align}
    I_1 &= \frac{1}{\mu} \int_{|\mu\omega| < \delta} \frac{d\omega}{2\pi} \ \tilde{\mathcal{D}}_{V}(\mu \omega) \left|\tilde{g}\left(\omega\right)\right|^2 \notag\\& = \frac{C_1}{\mu^{1+|\alpha|}} \int_{|\mu\omega| < \delta} \frac{d\omega}{2\pi} \ \frac{\left|\tilde{g}\left(\omega\right)\right|^2}{|\omega|^{|\alpha|}} \notag\\&= \frac{C_1}{\mu^{1+|\alpha|}} \int_{-\delta/\mu}^{\delta/\mu} \frac{d\omega}{2\pi} \ |\omega|^{2-|\alpha|}\left|\tilde{f}\left(\omega\right)\right|^2.
\end{align}
In the limit that $\mu\to 0$, the integral in the above expression becomes $\int_{-\infty}^{\infty} \frac{d\omega}{2\pi} \ |\omega|^{2-|\alpha|}\left|\tilde{f}\left(\omega\right)\right|^2$. This integral is finite, assuming that $\tilde{f}$ is well-behaved near $\omega=0$. Thus, $I_1$ scales as $1/\mu^{1+|\alpha|}$.

On the other hand, the second integral can again be written as
\begin{align}
    |I_2| &\leq \frac{1}{\mu} \int_{|\mu\omega| > \delta} \frac{d\omega}{2\pi} \ |\tilde{\mathcal{D}}_{V}(\mu \omega)| \left|\tilde{g}\left(\omega\right)\right|^2 \notag\\& = \frac{1}{\mu^2} \int_{|\omega| > \delta} \frac{d\omega}{2\pi} \ |\tilde{\mathcal{D}}_{V}(\omega)| \left|\tilde{g}\left(\frac{\omega}{\mu}\right)\right|^2 \notag\\&\leq \frac{C_2^2}{\pi} \int_{\delta}^\infty d\omega \ \frac{|\tilde{\mathcal{D}}_{V}(\omega)|}{\omega^2},
\end{align}
and therefore, is finite. Thus, in chaotic systems, the speed-Fisher information scales as $1/\mu^{1+|\alpha|}$.

The above argument also holds for marginally unstable systems, where $0 < \alpha < 1$. Repeating the same steps as above, one can show that the speed-Fisher information scales as $1/\mu^{1-|\alpha|}$ in the limit $\mu\to 0$.

\section{Relative Entropy Production}
\label{app_entropy}

Let us first consider a quantum system, and start from the definition of the relative entropy between the initial and final states, given in Eq.~(\ref{eq_dkl_defnQuantum}). The Dyson expansion allows us to compute this entropy up to leading order in $\bar{v}$:
\begin{align}
    &D_{\text{KL}}(\rho_{+\infty} || \rho_{-\infty}) \notag\\&= \frac{\bar{v}^2}{2\hbar^2\mu^2} \int_{-\infty}^\infty d\tau_1 \int_{-\infty}^\infty d\tau_2 \ f(\mu\tau_1) f(\mu\tau_2) \notag\\&\qquad\qquad\qquad\qquad\qquad\left\langle [V(\tau_1),[V(\tau_2),\ln\rho_{-\infty}]] \right\rangle.
\end{align}
Assuming $\rho_{-\infty} = P(H_0)$ to be a smooth function of the Hamiltonian, the correlation function in the above expression can be written in the energy eigenbasis as
\begin{align}
    &\left\langle [V(\tau_1),[V(\tau_2),\ln\rho_{-\infty}]] \right\rangle \notag\\&= \sum_{m\neq n} (P(E_m)-P(E_n))(\ln P(E_m) - \ln P(E_n)) \notag\\&\qquad\qquad\qquad\qquad\qquad\qquad\qquad|V_{mn}|^2 e^{i\omega_{mn}(\tau_1-\tau_2)}.
\end{align}
As before, we are interested in the low-frequency behavior of the spectral function, and thus, we only keep terms with $E_m\approx E_n$ in the above expression, giving us
\begin{align}
    &\left\langle [V(\tau_1),[V(\tau_2),\ln\rho_{-\infty}]] \right\rangle \notag\\&\simeq \hbar^2\sum_{m\neq n} P(E_m) (\partial_E \ln P(E_m))^2 \omega_{mn}^2 \notag\\&\qquad\qquad\qquad\qquad\qquad|V_{mn}|^2 e^{i\omega_{mn}(\tau_1-\tau_2)}.
\end{align}
And therefore, the relative entropy can be written in terms of the speed-Fisher information as
\begin{equation}
    D_{\text{KL}}(\rho_{+\infty} || \rho_{-\infty}) = \frac{1}{2}\mathscr{I}_{\bar{v}}(\mu)\bar{v}^2 + \mathcal{O}(\bar{v}^3).
\end{equation}
A similar argument can be repeated in the classical case to again recover the above expression.

From the definitions of the modular Hamiltonian and the Gibbs entropy, it is easy to show in both quantum and classical cases that
\begin{equation}
    D_{\text{KL}}(\rho_{+\infty} || \rho_{-\infty}) = \langle\Delta K_0\rangle - \Delta S.
\end{equation}
In Hamiltonian systems and volume-preserving autonomous systems, $\Delta S = 0$.

Focusing on quantum systems again, let us write the modular Hamiltonian in the energy eigenbasis as $K_0 = -\sum_n \ln P(E_n) \ket{n}\bra{n}$. We then define the modular energy difference matrix $\Delta K_0$ as
\begin{equation}
    (\Delta K_0)_{mn} = -\ln P(E_m) + \ln P(E_n).
\end{equation}
The probability of the system, when initially prepared in the state $\ket{n}$, transitioning to the state $\ket{m}$ at the end of the cycle is $\mathcal{P}_{n\to m} = |\bra{m}U_I(+\infty,-\infty)\ket{n}|^2$, where $U_I$ is the time evolution operator in the interaction picture. Then, it is easy to prove the generalized Jarzynski identity~\cite{jarzynski_1997}:
\begin{equation}
    \langle e^{-\Delta K_0}\rangle = \sum_{m,n} p_n \mathcal{P}_{n\to m} e^{\ln \frac{p_m}{p_n}} = 1.
\end{equation}
Taking the logarithm on both sides of the above expression, and using the cumulant expansion for $\ln \langle e^{-\Delta K_0}\rangle$, we find that
\begin{equation}
    -\langle \Delta K_0\rangle + \frac{1}{2}\text{Var}(\Delta K_0) + \dots = 0.
\end{equation}
If all other higher-order cumulants are negligible, one can replace the mean change in modular energy by its variance:
\begin{equation}
    \mathscr{I}_{\bar{v}}(\mu) \simeq \frac{\text{Var}(\Delta K_0)}{\bar{v}^2}.
\end{equation}
A similar argument can be made for classical Hamiltonian systems.

\begin{widetext}

\section{Exact Quantum Speed-Fisher Information}
\label{app_quantumScaling}

In a many-body quantum system with a Lorentzian spectral function, applying an exponentially modulated protocol produces the following speed-Fisher information:
\begin{equation}
    \mathscr{I}_{\bar{v}}(\mu) = \frac{8C_0\Gamma}{\hbar^2\mu^3} \int_{-\infty}^\infty \frac{d\omega}{2\pi} \frac{\tanh^2\left(\frac{\hbar\mu\omega}{2\kB T}\right)}{(\Gamma^2 + \mu^2\omega^2)(1 + \omega^2)^2},
\end{equation}
where $\Gamma = \nicefrac{1}{\tau_{\text{rel}}}$. This integral can subsequently be split into
\begin{align}
    \mathscr{I}_{\bar{v}}(\mu) =& \frac{8C_0\Gamma\mu}{\hbar^2(\Gamma^2-\mu^2)^2} \int_{-\infty}^\infty \frac{d\omega}{2\pi} \frac{\tanh^2\left(\frac{\hbar\mu\omega}{2\kB T}\right)}{\Gamma^2 + \mu^2\omega^2}
    - \frac{8C_0\Gamma}{\hbar^2\mu(\Gamma^2-\mu^2)^2} \int_{-\infty}^\infty \frac{d\omega}{2\pi} \frac{\tanh^2\left(\frac{\hbar\mu\omega}{2\kB T}\right)}{1 + \omega^2} \notag\\&
    + \frac{8C_0\Gamma}{\hbar^2\mu^3(\Gamma^2-\mu^2)} \int_{-\infty}^\infty \frac{d\omega}{2\pi} \frac{\tanh^2\left(\frac{\hbar\mu\omega}{2\kB T}\right)}{(1 + \omega^2)^2}.
\end{align}
We denote the three integral expressions by $I_1$, $I_2$, and $I_3$, with the sign convention $\mathscr{I}_{\bar v}(\mu)=I_1-I_2+I_3$.

First, focusing on the integral $I_1$, we express it as
\begin{equation}
    I_1 = \frac{4C_0}{\hbar^2(\Gamma^2 - \mu^2)^2}  - \frac{8C_0\Gamma\mu}{\hbar^2(\Gamma^2-\mu^2)^2} \int_{-\infty}^\infty \frac{d\omega}{2\pi} \frac{\text{sech}^2\left(\frac{\hbar\mu\omega}{2\kB T}\right)}{\Gamma^2 + \mu^2\omega^2}.
\end{equation}
The integral in the above expression can be evaluated by constructing a contour along the real line and an arc in the upper half of the complex plane. The contribution from the arc vanishes as its radius $R\to\infty$. Note that the integrand has poles at $\nicefrac{i\Gamma}{\mu}$ and at $\frac{2i\pi\kB T}{\hbar\mu}\left(n+\frac{1}{2}\right)$, for nonnegative integers $n$. Evaluating the residues at these poles, we find that
\begin{align}
    I_1 =& \frac{4C_0}{\hbar^2(\Gamma^2 - \mu^2)^2} - \frac{4 C_0\sec^2\left(\frac{\hbar\Gamma}{2\kB T}\right)}{\hbar^2(\Gamma^2-\mu^2)^2}
    + \frac{4C_0 }{\hbar^2 \pi^2(\Gamma^2-\mu^2)^2}\sum_{n=0}^\infty \left[\frac{1}{ ( n+\frac{1}{2} - \frac{\hbar\Gamma}{2\pi\kB T})^2}
    - \frac{1}{ (n + \frac{1}{2} + \frac{\hbar\Gamma}{2\pi\kB T})^2}\right] \notag\\ =&
    \frac{4C_0}{\hbar^2(\Gamma^2 - \mu^2)^2} - \frac{4 C_0\sec^2\left(\frac{\hbar\Gamma}{2\kB T}\right)}{\hbar^2(\Gamma^2-\mu^2)^2}
    + \frac{4C_0 }{\hbar^2 \pi^2 (\Gamma^2-\mu^2)^2} \left[ \psi^{(1)}\left(\frac{1}{2} - \frac{\hbar\Gamma}{2\pi\kB T}\right) 
    - \psi^{(1)}\left(\frac{1}{2} + \frac{\hbar\Gamma}{2\pi\kB T}\right)\right],
\end{align}
where $\psi^{(k)}(z)$ is the polygamma function, given by the series
\begin{equation}
    \psi^{(k)}(z) = (-1)^{k+1} k! \sum_{n=0}^\infty \frac{1}{(n+z)^{k+1}}.
\end{equation}
Using the reflection relation $\psi^{(1)}(1-z) + \psi^{(1)}(z) = \frac{\pi^2}{\sin^2\pi z}$, the above integral can be written as
\begin{align}
    I_1 =&
    \frac{4C_0}{\hbar^2(\Gamma^2 - \mu^2)^2}
    - \frac{8C_0 }{\hbar^2 \pi^2 (\Gamma^2-\mu^2)^2} \psi^{(1)}\left(\frac{1}{2} + \frac{\hbar\Gamma}{2\pi\kB T}\right).
\end{align}
In a similar fashion, one can compute $I_2$ and $I_3$ as:
\begin{align}
    I_2 =& \frac{8C_0\Gamma}{\hbar^2\mu(\Gamma^2-\mu^2)^2} \int_{-\infty}^\infty \frac{d\omega}{2\pi} \frac{\tanh^2\left(\frac{\hbar\mu\omega}{2\kB T}\right)}{1 + \omega^2} =
    \frac{4C_0\Gamma}{\hbar^2\mu(\Gamma^2-\mu^2)^2}
    - \frac{8 C_0 \Gamma}{\hbar^2\pi^2\mu(\Gamma^2-\mu^2)^2}\psi^{(1)}\left(\frac{1}{2} + \frac{\hbar\mu}{2\pi\kB T}\right),
\end{align}
and
\begin{align}
    I_3 &= \frac{8C_0\Gamma}{\hbar^2\mu^3(\Gamma^2-\mu^2)} \int_{-\infty}^\infty \frac{d\omega}{2\pi} \frac{\tanh^2\left(\frac{\hbar\mu\omega}{2\kB T}\right)}{(1 + \omega^2)^2} \notag\\=&
    \frac{2C_0\Gamma}{\hbar^2\mu^3(\Gamma^2-\mu^2)}
    - \frac{4 C_0\Gamma}{\hbar^2\mu^3\pi^2(\Gamma^2-\mu^2)} \psi^{(1)}\left(\frac{1}{2}+\frac{\hbar\mu}{2\pi\kB T}\right)
    + \frac{2 C_0\Gamma}{\hbar\mu^2\pi^3\kB T(\Gamma^2-\mu^2)} \psi^{(2)}\left(\frac{1}{2}+\frac{\hbar\mu}{2\pi\kB T}\right).
\end{align}
Combining all the above expressions together, we find the speed-Fisher information to be
\begin{align}
    \mathscr{I}_{\bar{v}}(\mu) = & \ \frac{2C_0(\Gamma + 2\mu)}{\hbar^2\mu^3(\Gamma+\mu)^2}
    - \frac{8C_0 }{\hbar^2 \pi^2 (\Gamma^2-\mu^2)^2} \psi^{(1)}\left(\frac{1}{2} + \frac{\hbar\Gamma}{2\pi\kB T}\right)
    - \frac{4C_0 \Gamma (\Gamma^2 - 3\mu^2)}{\hbar^2\pi^2\mu^3(\Gamma^2-\mu^2)^2}\psi^{(1)}\left(\frac{1}{2} + \frac{\hbar\mu}{2\pi\kB T}\right) \notag\\&
    + \frac{2 C_0\Gamma}{\hbar\mu^2\pi^3\kB T(\Gamma^2-\mu^2)} \psi^{(2)}\left(\frac{1}{2}+\frac{\hbar\mu}{2\pi\kB T}\right).
    \label{eq_exactQFI}
\end{align}

Note that this expression has a well-defined non-singular limit at $\Gamma\to \mu$. The classical limit is obtained by expanding the polygamma functions in the above expression in powers of $\hbar$ and then taking the limit $\hbar\to 0$: 
\begin{equation}
    \mathscr{I}_{\bar{v}}(\mu) = \frac{C_0\Gamma}{2\mu(\kB T)^2(\Gamma+\mu)^2}.
\end{equation}
On the other hand, if both $\frac{\hbar\Gamma}{\kB T} \gg 1$ and $\frac{\hbar\mu}{\kB T} \gg 1$ in the quantum regime, then
\begin{equation}
    \mathscr{I}_{\bar{v}}(\mu) \simeq \frac{2C_0(\Gamma + 2\mu)}{\hbar^2\mu^3(\Gamma+\mu)^2}.
\end{equation}

\end{widetext}

\bibliography{references}

\end{document}

%% file: standardPkgs.tex
\usepackage{amsmath}
\usepackage{amssymb}
\usepackage{braket}
\usepackage{bbm}
\usepackage{graphicx}
\usepackage{dcolumn}
\usepackage{bm}
\usepackage{tikz}
\usepackage{placeins}
\usepackage{nicefrac}
\usepackage{makecell}
\usepackage{caption}
\usepackage{tikz}
\usetikzlibrary{tikzmark,fit}
\usepackage{subcaption}
\usepackage{verbatim}
\usepackage[normalem]{ulem}
\usepackage{xurl}
\usepackage[dvipsnames]{xcolor}
\usepackage{array}
\usepackage{multirow}
\usepackage{booktabs}
\usepackage{comment}
\usepackage{amsthm}
\usepackage{mwe}
\usepackage{mathrsfs}
\allowdisplaybreaks

\usepackage{hyperref}
\hypersetup{
    colorlinks=true,
    linkcolor=blue,
    filecolor=blue,      
    urlcolor=blue,
    citecolor = blue,
    breaklinks=true,
}